\pdfoutput=1

\documentclass[11pt,nofootinbib]{article}
\usepackage{amsmath,amssymb,mathtools}
\usepackage{graphicx}
\usepackage[colorlinks=true,linkcolor=blue,citecolor=blue,urlcolor=blue,linktocpage=true]{hyperref}
\usepackage{subfigure}
\usepackage{comment}
\usepackage{tensor}
\usepackage{cases}
\usepackage[utf8]{inputenc}
\usepackage{cite}

\numberwithin{equation}{section}

\begin{document}

\providecommand{\abs}[1]{\lvert#1\rvert}
\providecommand{\bd}[1]{\boldsymbol{#1}}

\begin{titlepage}

\setcounter{page}{1} \baselineskip=15.5pt \thispagestyle{empty}

\begin{flushright}
% preprint number
\end{flushright}
\vfil

\bigskip
\begin{center}
 {\LARGE \textbf{Inflaxion Dark Matter}}
\vskip 15pt
\end{center}

\vspace{0.5cm}
\begin{center}
{\Large
Takeshi Kobayashi$^{\star,\dagger}$
and
Lorenzo Ubaldi$^{\ast,\dagger}$
}\end{center}

\vspace{0.3cm}

\begin{center}
\textit{$^{\star}$ International Centre for Theoretical Physics, Strada Costiera 11, 34151 Trieste, Italy}\\

\vskip 14pt 
\textit{$^{\ast}$ SISSA and INFN Sezione di Trieste, Via Bonomea 265, 34136 Trieste, Italy}\\

\vskip 14pt 
\textit{$^{\dagger}$ Institute for Fundamental Physics of the Universe,
 Via Beirut 2, 34014 Trieste, Italy}\\
 
\vskip 14pt
E-mail:
 \texttt{\href{mailto:takeshi@ictp.it}{takeshi@ictp.it}},
 \texttt{\href{mailto:ubaldi.physics@gmail.com}{ubaldi.physics@gmail.com}}
\end{center} 

%\vfil

%\vspace{0.8cm}

\vspace{1cm}

\noindent
A new mechanism for producing axion dark matter is proposed.
 By invoking low-scale inflation and a kinetic mixing between the axion
 and the inflaton, it is shown that the axion is driven
 to a field point slightly displaced from the potential minimum, which
 can give rise to the observed dark matter abundance.
 In this framework, different combinations of the axion and inflaton
 fields play various cosmological roles, including generating the
 cosmological perturbations, reheating the universe, and serving as dark
 matter. The kinetic mixing also relates the dark matter lifetime with
 the reheating temperature. The mechanism tames axions that would
 otherwise overdominate the universe, and thus opens up new windows in
 the axion parameter space, including decay constants at the GUT scale
 and higher. 
\vfil

\end{titlepage}

\newpage
\tableofcontents

\section{Introduction}
\label{sec:intro}

The remarkable agreement between measurements of the cosmic microwave background (CMB)
and the Lambda Cold Dark Matter ($\Lambda$CDM) model~\cite{Aghanim:2018eyx} gives us confidence that we understand
well several aspects of the early history of the universe. 
The existing data supports the picture that the Hot Big Bang cosmology
started with inflation, a period of exponentially accelerated
expansion~\cite{Starobinsky:1980te,Sato:1980yn,Guth:1980zm}. 
The simplest models of inflation involve a scalar
field slowly rolling down a flat potential. We know neither at what scale the process took place,
nor the temperature at which the plasma was reheated at the end of it. We are fairly certain that the reheat
temperature  must be above the MeV scale for the Big Bang Nucleosynthesis (BBN) to 
successfully take place. 
Most of the known mechanisms for generating the matter-antimatter
asymmetry require even higher reheating temperatures, 
however there are also baryogenesis models that can operate
at scales not much above BBN~\cite{Elor:2018twp}.
While (very) low-scale inflation remains as a possibility,
the improved observational limit on primordial
gravitational waves has tightened the bound on the inflation scale from
above, now disfavoring simple high-scale inflation 
models composed of single monomial terms~\cite{Akrami:2018odb}.

Another feature essential to our understanding of the universe is the existence of cold dark matter. Its actual
nature remains a mystery, but several 
candidates have been proposed. Among them, axions provide
a viable and well studied possibility;
these include the QCD axion which solves the strong $CP$
problem~\cite{Peccei:1977hh,Weinberg:1977ma,Wilczek:1977pj},  
as well as axion-like fields that can arise for instance from string theory
compactifications~\cite{Svrcek:2006yi,Douglas:2006es,Arvanitaki:2009fg}. 
Axions can have a wide range of masses and couplings to
matter, depending on the theoretical framework in which they are
embedded. 
Two properties of axions are of
particular importance in this study: (i) that they are pseudo
Nambu-Goldstone Bosons (pNGBs) of a spontaneously broken global
U(1)~symmetry, and as such obey
a shift symmetry, (ii) that an initial displacement of the axion field
from its potential minimum gives rise to a relic abundance of axion dark
matter in the later universe.

The most well-studied mechanism that gives rise to the axion 
displacement is the vacuum misalignment scenario, in which 
the symmetry breaking happens before or during the inflationary
epoch, and thus inflation renders the axion field homogeneous
within the observable universe.
The axion stays frozen at this initial field value while its mass 
is smaller than the Hubble expansion rate of the early universe,
then eventually starts to oscillate and behaves as cold dark matter.
For an axion-like field whose mass stays constant since the
inflationary epoch, the vacuum misalignment gives rise to a present-day
abundance of 
\begin{equation}
 \Omega_\sigma \sim 0.1 \times 
\left( \frac{\theta_{\star} f}{10^{17} \, \mathrm{GeV}} \right)^2
  \left( \frac{m_\sigma}{10^{-22}\, \mathrm{eV}} \right)^{1/2}.
\label{eq:axion-like-ab}
\end{equation}
Here $m_\sigma$ is the axion mass,
$f$ is the decay constant, 
and $\theta_\star$ is the misalignment angle
which parametrizes the initial field displacement 
as~$\sigma_\star = \theta_\star f$;
considering the axion potential to have a periodicity~$ \sigma \to
\sigma + 2 \pi f$, the angle lies within the range $-\pi < \theta_\star
< \pi$. 
An axion with $\abs{\theta_\star} \sim 1$
makes up the observed dark matter abundance if, for instance, 
the decay constant is $f \sim 10^{17}\, \mathrm{GeV}$
and the mass $m_\sigma \sim 10^{-22}\, \mathrm{eV}$.
This also implies that if $f \gtrsim 10^{17}\, \mathrm{GeV}$,
the axion would overdominate the universe and give rise to
the so-called cosmological moduli problem
unless $\theta_\star$ is sufficiently smaller than unity
and/or the mass satisfies
$m_\sigma \lesssim 10^{-22}\, \mathrm{eV}$.
An ultralight axion dark matter with $m_\sigma \sim 10^{-22}\,
\mathrm{eV}$ has been a subject of active
investigation over the years as it leaves distinct signatures in the
small-scale structures of the universe~\cite{Hu:2000ke}
(see e.g. \cite{Arvanitaki:2009fg,Hui:2016ltb} for reviews).
However it was recently pointed out that these are
in tension with observational data on the Lyman-$\alpha$
forest~\cite{Irsic:2017yje,Armengaud:2017nkf,Kobayashi:2017jcf}  
and galaxy rotation curves~\cite{Bar:2018acw}. 
In the vacuum misalignment scenario, 
the relevant dynamics for the axion field takes place after
inflation is over, and the abundance is derived by 
solving the equation of motion of the single axion field,
assuming that couplings to other fields can be neglected. 

In this paper we explore the possibility that the axion couples to the inflaton via a kinetic mixing term. 
From the point of view of
an effective field theory, with the axion and the inflaton being the
relevant degrees of freedom, 
there is no {\it a priori} reason to forbid such a mixing as it respects
the axion shift symmetry.
The main goal of this work is to demonstrate that this kinetic
mixing gives rise to a new mechanism for producing
axion dark matter.
We find the most interesting scenario to be 
when the Hubble scale of inflation, $H_{\rm inf}$, is smaller than the
axion mass,
so that the axion possesses a potential that noticeably affects its dynamics
during inflation.
In this case, without a kinetic mixing the
axion would simply be subject to damped oscillations, would relax to the minimum of its potential where its energy 
density vanishes, and thus would not contribute to the dark matter
abundance. 
We will show that the kinetic mixing keeps the axion displaced from its minimum during inflation, thus providing a source for misalignment.
This opens up new windows of the parameter space of axion dark matter, relevant to some experimental
searches.  
In the diagonal basis, the oscillating field that serves as
dark matter is a linear combination of the 
inflaton and axion. Hence, we dub it the {\it inflaxion}. 

Kinetic mixings among multiple axions have been considered in previous 
literature~\cite{Babu:1994id,Cicoli:2012sz,Higaki:2014qua,Cicoli:2017zbx,Agrawal:2017cmd}, but to our knowledge
not in the context of inflation. Our inflaton can also be axion-like, but this is not a requirement necessary
to the end of our proposed mechanism.  

The axion, being a pNGB, is expected to be light. 
As in this work we focus on inflationary Hubble scales lower than $m_\sigma$, we are led to consider low-scale models of inflation. 
In Refs.~\cite{Graham:2018jyp,Guth:2018hsa} the authors considered the QCD axion, without
any direct coupling to the inflaton, and studied the regime 
$m_\sigma < H_{\mathrm{inf}} < \Lambda_{\rm QCD}$. In such a case
the axion already has a potential during inflation, but is not oscillating;
classically the field slowly rolls toward its minimum, but its quantum fluctuations tend to
keep it displaced from the minimum. Given a long enough period of inflation, the two competing effects lead to
an equilibrium configuration which results in a probabilistic distribution for the misalignment angle, typically peaked
at values $\abs{\theta_\star} \ll 1$. In turn, this opens up the large
$f$ window. 
However for this mechanism to be predictive, 
it is important to reach equilibrium; this requires an extraordinarily long
period of inflation of $10^{20}$ $e$-folds or more,
which could pose a challenge for inflationary model
building.\footnote{For other interesting ideas on axion dark matter with
low-scale inflation, see e.g.\cite{Daido:2017wwb,Co:2018phi}.}
In our scenario, on the other hand, we consider even lower inflation
scales of $H_{\mathrm{inf}} < m_\sigma$. 
The field evolution will be dominated by purely classical
dynamics, and there will be no strict requirement on the number of inflationary $e$-folds.

Our mechanism can be applied to generic axions.
We discuss both the QCD axion and axion-like particles 
in Section~\ref{sec:mixing} where we present the basic construction of our
mechanism.
Then in Section~\ref{sec:mechanism} we describe how it can produce dark
matter and give rise to a consistent cosmological history. 
We explore the parameter space in Section~\ref{sec:parameter}; 
here, with the aim of 
concentrating on the salient features of the dynamics while avoiding
complications, we mainly focus on axion-like particles whose mass
stays constant throughout the cosmic evolution. 
Then we conclude with a discussion of directions for further research in
Section~\ref{sec:conc}.  
Some calculational details and possibilities with the QCD axion 
are left to the appendices.
In Appendix~\ref{sec:diagonal} we introduce a field basis that
diagonalizes the kinetic and mass terms of the inflaton-axion system.
In Appendix~\ref{sec:displacement} we explain
axion misalignment and slow-roll inflation in the diagonal basis. 
In Appendix~\ref{app:QCDinflaxion} we comment on the implications for
the QCD axion.

\section{Inflaton-Axion Mixing}
\label{sec:mixing}

\subsection{Axions Coupled to a Strong Sector}
\label{sec:QCD}

Let us start by considering an axion that is coupled to some strong
sector. As an example of such a field, we focus on the QCD axion.
Then our mechanism is described by the following simple model where the
axion~$\sigma$ is derivatively coupled to another real field~$\phi$, 
\begin{equation}
  \frac{\mathcal{L}}{\sqrt{-g} } =
-\frac{1}{2} g^{\mu \nu} \partial_\mu \sigma  \partial_\nu \sigma 
+ \frac{g_s^2}{32 \pi^2} \frac{\sigma }{f} G_{A \mu \nu} \tilde{G}_A^{\mu \nu} 
-\frac{1}{2} g^{\mu \nu} \partial_\mu \phi  \partial_\nu \phi 
- V(\phi)
- \alpha \, g^{\mu \nu} \partial_\mu \phi \partial_\nu \sigma ,
\label{eq:inflaton-QCD}
\end{equation}
where $f$ is the axion decay constant, $G_{A \mu \nu}$ the gluon
field strength.
We refer to $\phi$ as the inflaton, since we will be 
supposing that its effective potential~$V(\phi)$ has a shape capable of
inducing cosmic inflation as well as providing a graceful exit from it.
The axion field, odd under parity, respects a discrete shift
symmetry $\sigma \to \sigma + 2\pi f$,
which is not violated by the kinetic mixing term. 
By requiring the eigenvalues of the kinetic matrix to be positive 
(otherwise one of the physical degrees of freedom would be non-dynamical
or a ghost), the dimensionless coupling~$\alpha$ of the kinetic
mixing is restricted to lie within the range 
\begin{equation}
 -1 < \alpha < 1.
\label{eq:a-range}
\end{equation}

The kinetic terms can be diagonalized by redefining the fields,
for instance, as
\begin{equation}
 \sigma = \frac{1}{\sqrt{1-\alpha^2} } \, \Sigma,
\quad
 \phi = \Phi - \frac{\alpha }{\sqrt{1-\alpha^2}} \, \Sigma ,
\end{equation}
for which the Lagrangian takes the form
\begin{equation}
  \frac{\mathcal{L}}{\sqrt{-g} } =
-\frac{1}{2} g^{\mu \nu} \partial_\mu \Sigma  \partial_\nu \Sigma 
+ \frac{g_s^2}{32 \pi^2} \frac{\Sigma }{F}  G \tilde{G}
-\frac{1}{2} g^{\mu \nu} \partial_\mu \Phi  \partial_\nu \Phi 
- V\left( \Phi - \frac{\alpha }{\sqrt{1-\alpha^2}} \, \Sigma  \right).
\label{eq:Sigma-Phi}
\end{equation}
Ignoring for a moment the potential $V$, then $\Sigma$ is a standard canonical axion with a decay constant 
$F = \sqrt{1-\alpha^2} \, f$. 
The coupling to the gluons yields an effective potential for $\Sigma$
with periodicity~$2 \pi F$ which can roughly be approximated by a cosine,
\begin{equation}
U(\Sigma, T) = 
m_{\Sigma}^2 (T) \, F^2 
\left\{ 1 - \cos \left(\frac{\Sigma }{F} \right)  \right\}.
\label{eq:QCDpot}
\end{equation}
The potential vanishes at temperatures much higher than the QCD scale
$\Lambda_{\mathrm{QCD}} \approx 200\, \mathrm{MeV}$, 
while at temperatures below $\Lambda_{\mathrm{QCD}}$,
its mass at the minimum is~\cite{diCortona:2015ldu}
\begin{equation}
 m_{\Sigma}  \approx 6 \times 10^{-6}\, \mathrm{eV}
\left( \frac{10^{12}\, \mathrm{GeV}}{F} \right)
\quad
\mathrm{for}
\, \, \, 
T \ll \Lambda_{\mathrm{QCD}}.
\label{eq:QCDmass}
\end{equation}
Going back to the $(\sigma, \phi)$~basis, $\sigma$ similarly
obtains an effective potential, but with a periodicity~$2 \pi f$;
rewriting as $(\Sigma, F) \to (\sigma, f)$ in 
(\ref{eq:QCDpot}) and (\ref{eq:QCDmass})
gives the potential and mass for~$\sigma$. 

Now, let us consider the effects of $V$.
The shift symmetry of~$\sigma$, which was manifest in (\ref{eq:inflaton-QCD}), is realized in (\ref{eq:Sigma-Phi})
as a symmetry under a shift of both $\Sigma$ and $\Phi$ that keeps the
combination $\Phi - (\alpha /\sqrt{1-\alpha^2})\Sigma $ fixed. 
If we assume that $\phi$ is also parity-odd and that the minimum of $V$ preserves parity,
then our theory in the vacuum respects $P$ and $CP$. 
(The assumption of $CP$ conservation is not necessary for the dark matter
production mechanism we describe later, but is useful for 
illustrating the nature of the vacuum.)
Setting the minimum of $V(\phi)$ as $\phi_{\mathrm{min}} = 0$,
then the global minima of the full scalar potential at zero temperature,
$U(\Sigma, T = 0) + V$, 
are at the $CP$ conserving points 
\begin{equation}
 \Sigma_{\rm min} = 2\pi k F, 
 \quad
 \Phi_{\rm min} = \frac{\alpha}{\sqrt{1-\alpha^2}} 2\pi k F,
\end{equation}
with $k$ an integer.
In Minkowski spacetime, displacing the fields from a minimum results in coupled oscillations. 
The fields' values averaged over
a period longer than the timescale of the oscillations still correspond to $\Sigma_{\rm min}$ and $\Phi_{\rm min}$.
However in an expanding universe, given that $V$ is sufficiently flat,
the Hubble friction can keep the field $\Phi$ in slow-roll motion
while it approaches a minimum. When $\Phi$ is effectively frozen at a
point away from the minima, the axion~$\Sigma$ would also get displaced
from its $CP$-conserving minima.
The same effect is seen in the original basis~(\ref{eq:inflaton-QCD}) as 
$V$ rendering the field~$\phi$ spacetime-dependent and thus 
sourcing~$\square \phi$; this in turn generates an effective linear
potential for~$\sigma$ via the kinetic mixing.
We discuss these dynamics in detail in Section~\ref{sec:mechanism}.

We should also remark that from an effective field theory point of view, 
there can further be potential couplings of the form $I(\phi)
\cos(\sigma/f)$ which do not violate the axion's discrete shift symmetry.
Throughout this paper we assume such terms to be negligible.
However even if they were not, our mechanism of axion production can
work, but in parameter regions different from what we show in the
following discussions.

\subsection{General Axions}
\label{sec:general}

In the following, we will also be discussing axion-like fields which are
not necessarily coupled to any strong sector.
Hence hereafter we consider $\sigma$ to be a pNGB
of some global U(1)~symmetry that is explicitly broken (for instance by quantum
gravitational effects~\cite{Kallosh:1995hi, Alonso:2017avz}), 
and analyze a model of the form
\begin{equation}
  \frac{\mathcal{L}}{\sqrt{-g} } =
-\frac{1}{2} g^{\mu \nu} \partial_\mu \sigma  \partial_\nu \sigma 
- \frac{1}{2} m_\sigma^2 \sigma^2 
-\frac{1}{2} g^{\mu \nu} \partial_\mu \phi  \partial_\nu \phi 
- V(\phi) 
- \alpha \, g^{\mu \nu} \partial_\mu \phi \partial_\nu \sigma 
+ L_c[\sigma, \phi, \Psi].
\label{eq:Lagrangian}
\end{equation}
The mass term of the axion is understood to arise from expanding the
periodic potential around one of the minima, and the axion
mass~$m_\sigma$ may or may not depend on the temperature.
$L_c$ represents the couplings of $\sigma$ and $\phi$ to
other matter fields which we have collectively denoted as~$\Psi$;
its details will later be discussed.

We will see in the following sections that different combinations of
$\phi$ and $\sigma$ --- the inflaxions --- will play various
cosmological roles such as a canonical inflaton
producing the cosmological perturbations, a reheaton for heating up
the universe, and dark matter.

\section{Inflaxion Mechanism}
\label{sec:mechanism}

Let us now show how the model~(\ref{eq:Lagrangian}) 
produces dark matter, and gives rise to a consistent cosmological
history. 
We consider a flat FRW universe with metric
\begin{equation}
 ds^2 = -dt^2 + a(t)^2 d\bd{x}^2,
\label{eq:FRW}
\end{equation}
in which the homogeneous equations of motion of the fields read
\begin{align}
(1 - \alpha^2) (\ddot{\sigma} + 3 H \dot{\sigma})  &= 
\alpha V'(\phi)  - m_\sigma^2 \sigma,
\label{axionEOM}
\\
 (1 - \alpha^2)(\ddot{\phi} + 3 H \dot{\phi}) &= 
 - V'(\phi) + \alpha \, m_\sigma^2  \sigma.
\label{inflatonEOM}
\end{align}
Here we have supposed the axion mass~$m_\sigma$ to be a constant,
and ignored the matter coupling~$L_c$. 
An overdot represents a $t$-derivative,
$H = \dot{a} / a$ is the Hubble rate,
and $V'(\phi) = dV(\phi) / d\phi$.

\subsection{Low-Scale Inflation}
\label{sec:inf}

The inflaton potential~$V(\phi)$ is assumed to give rise to a
sufficiently long period of cosmic inflation, produce curvature
perturbations in good agreement with observation, 
and provide a graceful exit from inflation.\footnote{If the third and
higher order derivatives of~$V(\phi)$ are negligible, then one can
diagonalize the fields such that they do not interact with each other. Our
assumption for $V(\phi)$ can thus be rephrased as the requirement that
one of the diagonal fields possesses a flat enough potential to drive
inflation. See Appendix~\ref{sec:displacement} for detailed discussions.}
Our mechanism is insensitive to the detailed form of the inflaton potential,
however, we assume the energy scale of inflation to be sufficiently low
such that the U(1) symmetry is broken before inflation, and
also that the axion potential exists during inflation.
We identify the symmetry breaking scale with the axion decay constant~$f$, 
and in cases where the axion potential arises only below some energy
scale, we denote this by~$\Lambda$ (which is 
$\Lambda_{\mathrm{QCD}} $ for a QCD axion).
Then our requirements translate into a bound on the de Sitter temperature
during inflation,
\begin{equation}
 \frac{H_{\mathrm{inf}}}{2 \pi } < f, \Lambda ,
\label{eq:ab}
\end{equation}
where $H_{\mathrm{inf}}$ represents the typical value of the Hubble rate
during inflation.

Furthermore, we assume the inflationary scale to be even lower than the
axion mass, 
\begin{equation}
 H_{\mathrm{inf}} < m_\sigma. 
\label{eq:c}
\end{equation}
This forces the axion to undergo a damped oscillation along its
potential, and thus eventually be stabilized at the
minimum.\footnote{When the axion mass is not much larger but only
comparable to the Hubble scale, 
the axion would not oscillate but instead rapidly roll towards 
the minimum~\cite{Kofman:2007tr,Kobayashi:2009nv}.
Hence also in this case the axion approaches the
solution~(\ref{eq:static}).} 
However, with the kinetic mixing with the inflaton, 
the axion's effective potential minimum is shifted from the origin.
This is seen by setting the left hand side of~(\ref{axionEOM}) to zero
to obtain the quasi-static solution, 
\begin{equation}
 \sigma \simeq \frac{\alpha V'(\phi)}{m_\sigma^2} 
 \simeq - \frac{3 \alpha H \dot{\phi}}{m_\sigma^2}.
\label{eq:static}
\end{equation}
Upon moving to the far right hand side, we have 
assumed the inflaton potential to satisfy the slow-roll conditions 
and used $3 H \dot{\phi} \simeq -V'$.
A more detailed derivation of this solution, together with the
verification of the slow-roll approximation for the
non-canonical~$\phi$,  are
given in Appendix~\ref{sec:displacement}. 
Because the axion settles down to the
field value~(\ref{eq:static}) during inflation,
the initial axion misalignment for the post-inflationary evolution is
uniquely fixed by the Lagrangian parameters,
unlike the case for the vacuum misalignment scenario.

Let us set the minimum of~$V(\phi)$ where the inflaton settles down to
after the end of inflation as $\phi = 0$.
It is seen from (\ref{eq:static}) that the field displacements during
inflation take hierarchical values,
\begin{equation}
 \left| \frac{\sigma }{\alpha \phi } \right| \simeq 
\left| -3 \frac{H^2}{m_\sigma^2} 
\frac{\dot{\phi}}{H \phi } \right|
\ll 1 ,
\label{eq:2.9}
\end{equation}
where the inequality follows from (\ref{eq:c})
and $\abs{\dot{\phi}/ H \phi } \ll 1$ (otherwise inflation would
quickly end within a Hubble time).

With the condition~(\ref{eq:c}), inflation is effectively a single-field
model. (The canonical inflaton field whose fluctuation sources curvature
perturbations is given by a linear combination of $\sigma$ and $\phi$;
this is shown as $\varphi_{\mathrm{L}}$ in (\ref{varphi_HL}).)
Therefore no isocurvature fluctuations of dark matter is produced. 
We also remark that due to the low inflation scale, the inflationary
mechanism is envisaged to be a small-field model with tiny variation of
the Hubble rate during inflation.\footnote{However, in the presence of
spectator fields that predominantly produce the curvature perturbations
(such as curvatons), large-field models can also account for low
inflation scales, see e.g.~\cite{Kobayashi:2013bna}.}

\subsection{End of Inflation}
\label{sec:end}

\begin{figure}[t]
  \begin{center}
  \begin{center}
  \includegraphics[width=0.45\linewidth]{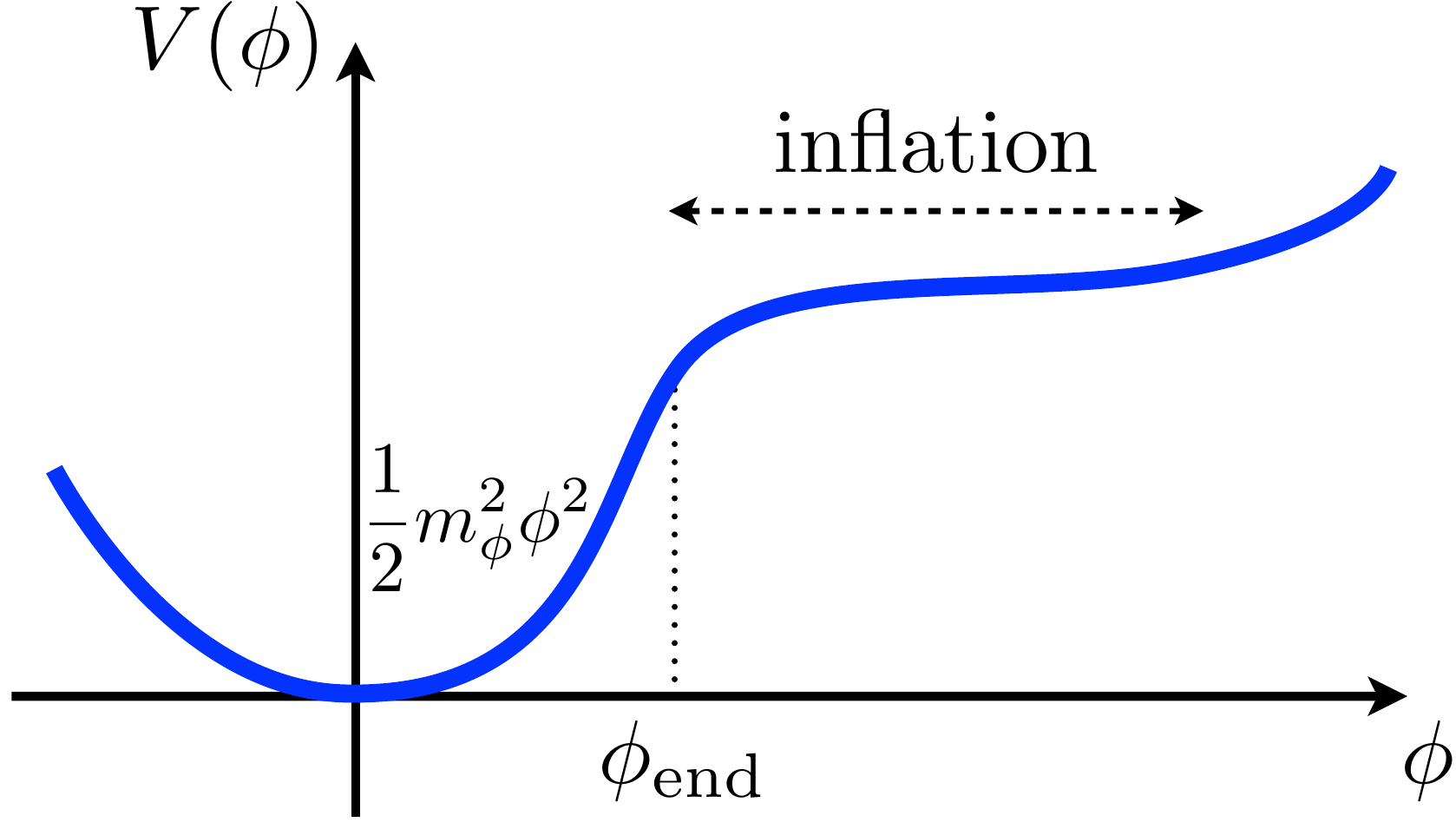}
  \end{center}
  \caption{Schematic of inflaton potential for small-field inflation.}
  \label{fig:V}
  \end{center}
\end{figure}

Slow-roll inflation comes to an end when the $\phi$~field approaches a
point along the potential where it is no longer flat. 
The field then rolls rapidly towards its potential minimum
at $\phi = 0$, and starts to oscillate. 
We suppose that the potential around the origin 
is approximately quadratic, 
up to the field value~$\phi_{\mathrm{end}}$ where inflation ends:
\begin{equation}
 V(\phi) \simeq \frac{1}{2} m_\phi^2 \phi^2
\quad
\mathrm{for}
\quad
\abs{\phi} \lesssim \abs{\phi_{\mathrm{end}}}.
\end{equation}
A schematic of the inflaton potential we are envisaging can be seen in Figure~\ref{fig:V}.
Then since the end of inflation
corresponds to the time when
$ V \simeq 2 M_p^2 H^2$ is satisfied
(cf.~(\ref{eq:end-V})), 
the inflaton field value at this point is obtained as
\begin{equation}
 \phi_{\mathrm{end}}^2 \simeq 
\left( \frac{2 M_p H_{\mathrm{end}}}{m_\phi } \right)^2,
\label{eq:phi_end}
\end{equation}
with $H_{\mathrm{end}}$ denoting the Hubble rate at the end of
inflation.

During inflation, the inflaton's effective mass is smaller than
the Hubble rate, and hence also smaller than the axion mass, 
$\abs{V''} \ll H_{\mathrm{inf}}^2 < m_\sigma^2$.
However the mass~$m_{\phi}$ at the vacuum can be much 
larger than $\abs{V''}^{1/2}$ during inflation;
this is particularly so for small-field inflation models in which 
various potential terms conspire 
to realize a flat potential in some field range.
It will turn out that, in order for the inflaton-axion system to provide
a dark matter candidate, the inflaton mass after inflation needs to be
larger than that of the axion, i.e.,
\begin{equation}
 m_\phi^2 \gg m_\sigma^2. 
\label{eq:e}
\end{equation}
Let us thus impose this condition hereafter.

In order to analyze the field dynamics in the post-inflationary epoch,
let us redefine the fields such that the kinetic and mass terms in the
Lagrangian are diagonalized, 
\begin{equation}
  \frac{\mathcal{L}}{\sqrt{-g} } = 
\sum_{i=\mathrm{DM}, \mathrm{RH}} \left(
-\frac{1}{2} g^{\mu \nu} \partial_\mu \varphi_i \, \partial_\nu \varphi_i
- \frac{1}{2} m_i^2 \varphi_i^2 
\right) + L_c[\sigma, \phi, \Psi].
\end{equation}
The exact expressions for the diagonal fields and their masses are given
in Appendix~\ref{sec:diagonal}. Here, under the mass
hierarchy~(\ref{eq:e}), the expressions at the leading order in
$m_\sigma^2 / m_\phi^2$ take a simple form of
\begin{gather}
\varphi_{\mathrm{DM}} \simeq
\alpha \, \phi + \sigma,
\quad
\varphi_{\mathrm{RH}} \simeq
\sqrt{1 - \alpha^2}
\left( \phi  - \alpha  \frac{m_\sigma^2}{m_{\phi}^2} \, \sigma \right),
\label{eq:214}
\\
 m_{\mathrm{DM}}^2 \simeq  m_\sigma^2,
\quad
 m_{\mathrm{RH}}^2 \simeq \frac{ m_{\phi}^2}{1-\alpha^2},
\label{eq:215}
\end{gather}
where in (\ref{eq:214}) we have displayed each of the coefficients of
$\phi$ and $\sigma$ at the leading order.
Each of the fields 
$\varphi_{\mathrm{DM}} $ and $\varphi_{\mathrm{RH}} $ 
undergoes harmonic oscillations about the origin, and thus behaves as
pressureless matter in the post-inflationary epoch.
$\varphi_{\mathrm{DM}}$ has approximately the same mass as the
axion, and will serve as a dark matter candidate, as its subscript
indicates. $\varphi_{\mathrm{RH}}$, on the other hand, is much heavier
(note that from (\ref{eq:a-range}) and (\ref{eq:e}) follows
$m_{\mathrm{DM}}^2 \ll m_{\mathrm{RH}}^2$), 
and this field will reheat the universe via its decay.

We defer the discussions on decay to the next section, and here
we estimate the energy densities of the oscillating fields.
As we saw in~(\ref{eq:2.9}), the field values of $\phi$ and $\sigma$
obey a hierarchy $\abs{\alpha \phi} \gg \abs{\sigma}$
during inflation.
This can break down towards the end of inflation, 
depending on the details of the exit from inflation.
However let us suppose that at the end of inflation the inequality is
still marginally satisfied, i.e., $\abs{\alpha \phi_{\mathrm{end}}}
\gtrsim \abs{\sigma_{\mathrm{end}}}$, 
and write the field displacement of 
$ \varphi_{\mathrm{DM}}$ using (\ref{eq:phi_end}) as
\begin{equation}
 \varphi_{\mathrm{DM} \, \mathrm{end}}^2 = 
\left( \frac{C \alpha M_p H_{\mathrm{end}}}{m_\phi}  \right)^2.
\label{eq:varphi_DMend}
\end{equation}
Here $C$ is a model-dependent dimensionless parameter
that is typically larger than unity due to the factor~$2$ in
(\ref{eq:phi_end}) and the contribution from $\sigma_{\mathrm{end}}$,
but not much larger, say, $C \sim 10$. 
Considering $ \varphi_{\mathrm{DM}}$ to start oscillating with this
initial amplitude at the end of inflation,
the ratio between its energy density 
which is initially dominated by the potential energy,
$ \rho_{\mathrm{DM}} \simeq m_{\mathrm{DM}}^2  \varphi_{\mathrm{DM}}^2
/2$, and the total density of the universe
$\rho_{\mathrm{tot}} 
= \rho_{\mathrm{DM}} + \rho_{\mathrm{RH}}
= 3 M_p^2 H^2$
is obtained as
\begin{equation}
  \frac{\rho_{\mathrm{DM}}}{\rho_{\mathrm{tot}}}
\simeq \frac{C^2 \alpha^2}{6} 
\frac{m_\sigma^2}{m_\phi^2}.
\label{eq:f}
\end{equation}
This density ratio stays constant while the fields oscillate
harmonically,\footnote{Depending on the inflaton
potential at around the end of inflation, the fields may 
exchange energy through a parametric resonance; 
then the dark matter fraction might be modified from the
estimate of~(\ref{eq:f}).} since both 
$\rho_{\mathrm{DM}}$ and $\rho_{\mathrm{RH}}$
redshift as $\propto a^{-3}$. 
One sees that the density fraction of $\varphi_{\mathrm{DM}}$ is
suppressed by the mass hierarchy~(\ref{eq:e}),
and thus the universe is dominated by $\varphi_{\mathrm{RH}}$, 
i.e., $\rho_{\mathrm{DM}} \ll \rho_{\mathrm{RH}} \simeq
\rho_{\mathrm{tot}}$.

\begin{figure}[t]
  \begin{center}
  \begin{center}
  \includegraphics[width=0.5\linewidth]{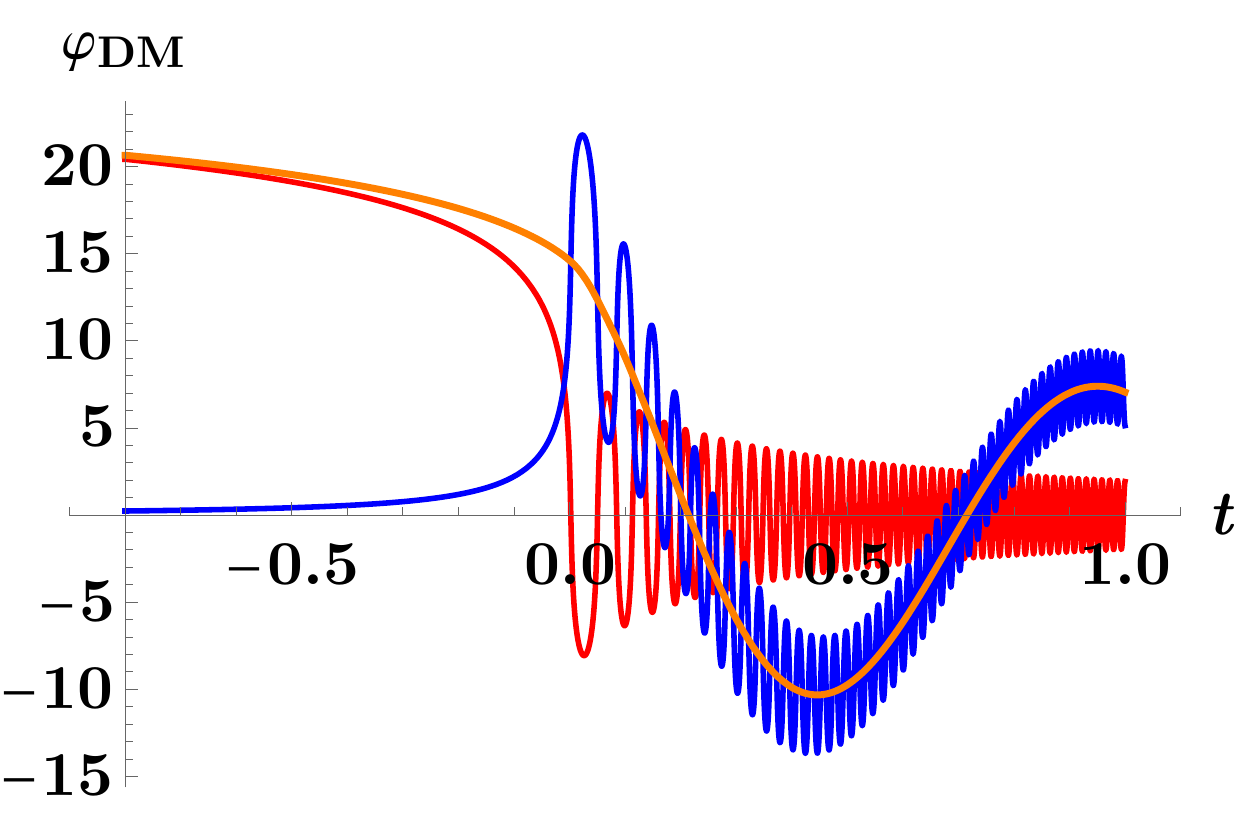}
  \end{center}
  \caption{Field evolution at the end of inflation for $\varphi_{\mathrm{DM}}$ (orange), $\sigma$ (blue), and $\alpha \phi$ (red). 
   Field values are normalized by $\alpha M_p H_{\mathrm{end}} / m_\phi$. 
   Time is in units of $2 \pi / m_\sigma$,
   with $t=0$ denoting when $-\dot{H} / H^2 = 1$.
   Shown is the case of
   $V(\phi) \propto 1- \mathrm{sech} (\phi/\mu) $,
   with $m_\sigma \approx 10 H_{\mathrm{inf}}$ and $m_\phi \approx 10^3
   H_{\mathrm{inf}}$.}
  \label{fig:dynamics}
  \end{center}
\end{figure}

The actual field evolutions at around the end of inflation can be seen in
Figure~\ref{fig:dynamics}.
For this plot we numerically solved the equations of motion
(\ref{axionEOM}) and (\ref{inflatonEOM})
together with the Friedmann equation,
for a case where the inflaton potential is of the form\footnote{This
potential is adopted as a toy model for describing the late stage of 
inflation and the subsequent oscillatory phase, hence we do
not worry whether it produces observationally viable curvature
perturbations on scales that have exited the Hubble horizon long before
the end of inflation.}
\begin{equation}
 V(\phi) = m_\phi^2 \mu^2 
\left( 1 - \frac{2}{e^{\phi/\mu} + e^{-\phi/\mu}}  \right).
\label{eq:sech}
\end{equation}
Shown are time evolutions of $\varphi_{\mathrm{DM}} = \alpha
\phi + \sigma$ (orange curve),
$\sigma$ (blue), and $\alpha \phi$ (red), with the
field values normalized by $\alpha M_p H_{\mathrm{end}} / m_\phi$. 
The horizontal axis
shows the dimensionless time in units of~$2 \pi / m_\sigma$ (i.e. the
oscillation period of $\varphi_{\mathrm{DM}}$),
and $t = 0$ is set to the end of inflation when $-\dot{H}/H^2 $ becomes 
unity. 
The parameters are chosen as
$\mu = 4 \times 10^{15} \, \mathrm{GeV}$, $m_\phi = 10^2 m_\sigma$, and
$\abs{\alpha} \ll 1$.
These conditions are enough to specify the field dynamics in terms of
the normalized quantities; in particular, the plot is independent of the
explicit values of $m_\phi$ and $\alpha$.
The inflation scale for the potential~(\ref{eq:sech})
is $H_{\mathrm{inf}} \simeq m_\phi \mu / \sqrt{3} M_p$,
hence our choice of~$\mu$ gives $m_\phi \approx 10^3 H_{\mathrm{inf}}$.
One clearly sees through this example that the
field~$\varphi_{\mathrm{DM}}$ starts oscillating at around the end of
inflation\footnote{Although the field $\varphi_{\mathrm{DM}} = \alpha
\phi + \sigma$ can be  defined at all times, one should keep in mind
that this forms a diagonal basis together with $\varphi_{\mathrm{RH}}$ 
only after inflation when the potential~(\ref{eq:sech}) has become
approximately quadratic.} 
with a frequency~$\sim m_\sigma$,
and an initial amplitude~(\ref{eq:varphi_DMend}) whose numerical
coefficient~$C$ is of order~$10$. 
The non-diagonal fields $\sigma$ and $\phi$
further carry an oscillatory mode with frequency~$\sim
m_{\phi}/\sqrt{1-\alpha^2}$, which corresponds to the
$\varphi_{\mathrm{RH}}$~component.

As we will show in the next section, 
the couplings of $\phi$ and $\sigma$ with other matter fields
provide decay channels for both diagonal fields, 
and induce decay rates that are typically larger for 
$\varphi_{\mathrm{RH}}$ than $\varphi_{\mathrm{DM}}$.
This allows $\varphi_{\mathrm{RH}}$ to decay and start the Hot Big Bang
cosmology, while $\varphi_{\mathrm{DM}}$ being long-lived.
For this scenario, it is essential that the condition (\ref{eq:e}) holds
so that the mass hierarchy between the inflaton and the axion flips
at the end of inflation, i.e.,
\begin{equation}
 \abs{V''} \ll m_\sigma^2 \quad (\mathrm{during\, \, inflation})
\quad \longrightarrow \quad
 m_\phi^2 \gg m_\sigma^2 \quad(\mathrm{after\, \, inflation}).
\end{equation}
Consequently the diagonal basis rotates in the
$\phi$ -- $\sigma$~plane (from (\ref{varphi_HL}) to (\ref{eq:214})),
giving most of the energy density in the post-inflation universe to the
heavier diagonal field.

If (\ref{eq:e}) does not hold and the axion mass remains larger, 
then the heavier diagonal field
(i.e. $\varphi_{\mathrm{H}}$ in (\ref{eq:sggp}))
would continue to be placed close to~$0$ after inflation,
carrying only a tiny fraction of the total energy density.
Reheating would then complete with the decay of the lighter 
field, which is typically also the longer-lived one,
and thus neither of the fields could survive until
today to serve as dark matter.  

Before ending this section we should remark that,
although we have been treating the axion potential as a quadratic,
it is actually periodic in~$\sigma$ with a periodicity 
of order the decay constant.
The quadratic approximation is valid in the vicinity of the potential
minima, and thus we have been implicitly assuming the axion displacement
to be of $ \abs{\sigma} < f$. 
Let us rewrite this condition at the end of inflation in a more
conservative form of
\begin{equation}
 \abs{\varphi_{\mathrm{DM} \, \mathrm{end}}} < f,
\label{eq:dp}
\end{equation}
and impose this as a constraint on the model parameters.
Violation of this condition implies that the
axion may rotate along its periodic potential multiple times during
inflation; the resulting random axion misalignment angle at the end of
inflation may also be able to explain the dark matter
abundance, however, we will not consider this possibility in the
following discussions.

\subsection{Reheating and Dark Matter Stability}
\label{sec:reh}

The axion~$\sigma$ can couple to other matter fields through
shift-symmetric operators.
Together with $\phi$'s coupling terms, they provide decay
channels\footnote{The phenomenology of kinetically mixed
axions interacting with other matter fields was also studied in
e.g. \cite{Higaki:2014qua,Agrawal:2017cmd}.} for
both $\varphi_{\mathrm{RH}}$ and
$\varphi_{\mathrm{DM}}$.
This is easily seen from inverting~(\ref{eq:214}), which at the leading
order in~$m_\sigma^2/ m_\phi^2$ gives
\begin{equation}
\sigma \simeq \varphi_{\mathrm{DM}} - \frac{\alpha }{\sqrt{1-
 \alpha^2}}\, \varphi_{\mathrm{RH}},
\quad
\phi \simeq
\alpha  \frac{m_\sigma^2}{m_{\phi}^2} \, \varphi_{\mathrm{DM}}
+ \frac{1}{\sqrt{1-\alpha^2}} \, \varphi_{\mathrm{RH}}.
\label{eq:318}
\end{equation}
Here one particularly sees that the couplings the
dark matter field~$\varphi_{\mathrm{DM}}$ obtains through $\phi$ is 
suppressed by~$(m_\sigma/m_\phi)^2$.

For concreteness, let us for example study the following matter couplings:
\begin{equation}
 L_c[\sigma, \phi, \Psi] = 
\frac{G_{\sigma \gamma \gamma }}{4}
\sigma F_{\mu \nu} \tilde{F}^{\mu \nu} 
 +  \frac{G_{\phi \gamma \gamma }}{4}
\phi F_{\mu \nu} \tilde{F}^{\mu \nu} 
+ g_{\phi ff} \, \phi \bar{\psi} i \gamma^5 \psi,
\label{eq:Lc3}
\end{equation}
where $F_{\mu\nu}$ is the field strength of either photons or hidden
photons, $\psi$ is a Dirac fermion, and 
$G_{\phi (\sigma) \gamma \gamma}$, $g_{\phi ff}$ are coupling constants. 
Here we have considered $\phi$ to be a pseudoscalar so that its kinetic
mixing term with the axion conserves parity.
As we want to reheat the visible sector above the BBN temperature, the obvious choice
is to couple the scalar fields to the visible photons and/or the Standard Model (SM) fermions. 
In this case, given the SM gauge symmetry, the coupling $g_{\phi ff}$ would contain the suppression
$m_\psi / M$, with $M$ some cutoff scale. 
If instead we reheated into a hidden sector, then we should address how to subsequently reheat the visible one.
That requires more model building, which is beyond the scope of this
paper. Here we only consider perturbative reheating,
for the sake of keeping the argument as concise and clear as possible. Another option is to study
preheating, which could be very efficient and would be of particular interest in the case of the QCD inflaxion. 
We leave such a study to future work.

The decay rates of the dark matter and reheaton fields to photons
via $\sigma F\tilde{F}$ are (here we neglect the possibility that the
contributions to the photon couplings through $\sigma$ and $\phi$
cancel each other)
\begin{equation}
 \Gamma (\varphi_\mathrm{DM} \to \gamma \gamma)
\simeq \frac{G_{\sigma \gamma \gamma }^2 }{64 \pi } m_\sigma^3,
\quad
 \Gamma (\varphi_\mathrm{RH} \to \gamma \gamma )
\simeq 
\frac{\alpha^2}{(1-\alpha^2)^{5/2}}
\frac{G_{\sigma \gamma \gamma }^2 }{64 \pi } m_\phi^3,
\label{eq:sgg}
\end{equation}
where we have written the diagonal masses in terms of $m_\sigma$ and
$m_\phi$ using~(\ref{eq:215}).
Here, the ratio between the decay rates of $\varphi_{\mathrm{DM}}$
and $\varphi_{\mathrm{RH}}$ is suppressed by the mass hierarchy as
$ \Gamma_{\mathrm{DM} }/  \Gamma_{\mathrm{RH} } \propto (m_\sigma / m_\phi)^3$. 
On the other hand, the two-photon decay rates induced by $\phi
F\tilde{F}$ are 
\begin{equation}
 \Gamma (\varphi_\mathrm{DM}  \to \gamma \gamma )
\simeq 
\alpha^2 
\frac{G_{\phi \gamma \gamma }^2 }{64 \pi }
\frac{m_\sigma^7}{m_\phi^4},
\quad
 \Gamma (\varphi_\mathrm{RH} \to \gamma \gamma )
\simeq 
\frac{1}{(1-\alpha^2)^{5/2}}
\frac{G_{\phi \gamma \gamma }^2 }{64 \pi } m_\phi^3,
\label{eq:pgg}
\end{equation}
with ratio 
$ \Gamma_{\mathrm{DM} }/  \Gamma_{\mathrm{RH} } \propto (m_\sigma / m_\phi)^7$.
The decay rates to fermions, if they are much lighter than
both scalars ($m_\psi \ll m_\sigma \ll m_\phi$) are
\begin{equation}
  \Gamma (\varphi_\mathrm{DM} \to f \bar{f} )
\simeq \alpha^2 \frac{g_{\phi ff}^2}{8 \pi }
\frac{m_\sigma^5}{m_\phi^4},
\quad
  \Gamma (\varphi_\mathrm{RH} \to f \bar{f} )
\simeq \frac{1}{(1-\alpha^2)^{3/2}} \frac{g_{\phi ff}^2}{8 \pi }
m_\phi,
\label{eq:pff}
\end{equation}
with ratio $ \Gamma_{\mathrm{DM} }/  \Gamma_{\mathrm{RH} } \propto (m_\sigma / m_\phi)^5$.
It should be noted that for the channels (\ref{eq:pgg}) and
(\ref{eq:pff}) induced by the inflaton~$\phi$, 
the decay rate of $\varphi_{\mathrm{DM}}$ is 
suppressed compared to that of $\varphi_{\mathrm{RH}}$
not only due to kinematics, but further because of the
suppression of the effective coupling as seen in~(\ref{eq:318}).
One can also envisage a situation where the coupled fermions are lighter
than the inflaton but heavier than the axion,
in which case the decay of $\varphi_{\mathrm{DM}}$ would be
kinematically forbidden.

Thus we have seen through the examples that,
unless $\alpha$ is tiny,
$\varphi_{\mathrm{DM}}$ is generically more stable than
$\varphi_{\mathrm{RH}}$, which allows for the possibility of
$\varphi_{\mathrm{DM}}$ surviving until today, while 
$\varphi_{\mathrm{RH}}$ quickly decaying and reheating the universe.
The time of reheating can be estimated as when 
the total decay rate~$\Gamma_{\mathrm{RH}}$ 
becomes comparable to the Hubble rate. 
If $\Gamma_{\mathrm{RH}}$ computed as above exceeds the Hubble
rate at the end of inflation, then $\varphi_{\mathrm{RH}}$ would 
quickly decay after inflation ends.\footnote{One should remember that 
the inflaton mass during inflation was much smaller than
$m_\phi$, and that $(\varphi_{\mathrm{DM}}, \varphi_{\mathrm{RH}})$ form a
diagonal basis only after inflation. Hence 
$\Gamma_{\mathrm{RH}} > H_{\mathrm{end}}$ does not necessary 
mean that inflaton fluctuations should have decayed during inflation.
The stabilized field (i.e. $\varphi_{\mathrm{H}}$ in
Appendix~\ref{sec:displacement}) could 
decay during inflation;
this would further damp the field's
oscillation and thus lead to a very quick realization of the 
quasi-static solution~(\ref{eq:static}).\label{fn:1}}
Thus the Hubble rate at reheating can be estimated as
\begin{equation}
 H_{\mathrm{reh}} = 
 \begin{dcases}
    H_{\mathrm{end}} 
       & \mathrm{for}\, \, \, \Gamma_{\mathrm{RH}}  >   H_{\mathrm{end}}, \\
    \Gamma_{\mathrm{RH}}
       & \mathrm{for}\, \, \, \Gamma_{\mathrm{RH}} \leq H_{\mathrm{end}}.
 \end{dcases}
\label{eq:g}
\end{equation}
The decay of $\varphi_{\mathrm{RH}}$ opens the radiation-dominated era, 
and assuming the decay products to quickly thermalize, the initial
temperature upon reheating is computed via
\begin{equation}
 3 M_p^2 H_{\mathrm{reh}}^2 = \frac{\pi^2}{30}g_* (T_{\mathrm{reh}}) \, 
T_{\mathrm{reh}}^4.
\label{eq:rho-T}
\end{equation}
This temperature is required from BBN to be at
least of~\cite{Kawasaki:2000en,Hannestad:2004px} 
\begin{equation}
 T_{\mathrm{reh}} \gtrsim 4 \, \mathrm{MeV},
\label{eq:h}
\end{equation}
which in turn sets a lower bound on~$\Gamma_{\mathrm{RH}}$.

As for $\varphi_{\mathrm{DM}}$, we require its lifetime to be longer
than the age of the universe,
\begin{equation}
 \Gamma_{\mathrm{DM}} < H_0,
\label{eq:j}
\end{equation}
where the Hubble constant in the current universe is 
$H_0 \approx 1 \times 10^{-33}\, \mathrm{eV}$~\cite{Aghanim:2018eyx}.
In order to compute its present-day abundance, let us make some 
simplifying assumptions about the reheating temperature.
Firstly, we assume it not to exceed
the symmetry breaking scale so that the 
symmetry stays broken and the 
axion exists throughout the post-inflationary epoch,
\begin{equation}
 T_{\mathrm{reh}} < f.
\label{eq:i}
\end{equation}
Moreover, in cases where the axion mass arises only below some
scale~$\Lambda$, we assume the reheating temperature to be also lower
than this, 
\begin{equation}
 T_{\mathrm{reh}} < \Lambda.
\label{eq:A}
\end{equation}
Under these simplifying assumptions, the energy density of 
$\varphi_{\mathrm{DM}}$ continues to redshift as 
$\rho_{\mathrm{DM}} \propto a^{-3}$ after reheating.
Assuming the entropy of the universe to be conserved after reheating,
the dark matter densities today and upon reheating can be related as
$\rho_{\mathrm{DM}\, 0} = \rho_{\mathrm{DM}\, \mathrm{reh}}
(s_0/s_{\mathrm{reh}})$ using the entropy density~$s$, which is a
function of the temperature as
\begin{equation}
 s = \frac{2 \pi^2}{45}g_{s*}(T)\,  T^3.
\label{eq:s-T}
\end{equation}
Combining these with (\ref{eq:f}) and (\ref{eq:rho-T}) yields 
the ratio of the dark matter density to the total density in the present
universe, 
\begin{equation}
 \Omega_{\mathrm{DM}} \simeq \frac{C^2 \alpha^2}{6} 
\frac{m_\sigma^2}{m_\phi^2} \frac{H_{\mathrm{reh}}^2 }{H_0^2}
\frac{s_0}{s_{\mathrm{reh}}}
\sim 
10^{-2} \, 
\alpha^2 \frac{m_\sigma^2}{m_\phi^2}
\left( \frac{H_{\mathrm{reh}}}{H_0} \right)^{1/2},
\label{eq:A2}
\end{equation}
where in the far right hand side we have substituted $C \sim 10$.
We also note that the detailed values of the
number of relativistic degrees of freedom~$g_{(s)*}$ upon reheating do
not affect the order-of-magnitude estimate.

Before turning to an analysis of the inflaxion parameter space,
let us mention that even if the conditions (\ref{eq:i}) and (\ref{eq:A})
are violated, we may still get dark matter.
Violation of~(\ref{eq:i}) implies there would be a second symmetry
breaking after reheating; this would lead to generation of topological
defects which may eventually emit inflaxion dark matter.
On the other hand by violating (\ref{eq:A}), 
the axion mass temporarily disappears upon reheating until the cosmic
temperature cools down again to~$\Lambda$; it would be interesting to
analyze whether the observed dark matter abundance can also be achieved
in this case.

\section{Parameter Space}
\label{sec:parameter}

The conditions we have imposed on the inflaxion parameters are 
(\ref{eq:ab}), (\ref{eq:c}), (\ref{eq:e}), (\ref{eq:dp}), (\ref{eq:h}),
(\ref{eq:j}), (\ref{eq:i}), and (\ref{eq:A}).
Furthermore, for $\varphi_{\mathrm{DM}}$ to make up the dark matter
of the universe, its abundance~(\ref{eq:A2}) should satisfy
$\Omega_{\mathrm{DM}} \approx 0.3$~\cite{Aghanim:2018eyx}.
Although some of the conditions were introduced only to simplify the
computation and thus not necessarily required for our mechanism to
provide dark matter, below we explore the parameter space that satisfies
all these conditions. 

Requiring $\varphi_{\mathrm{RH}}$ to quickly decay while
$\varphi_{\mathrm{DM}}$ to be long-lived imposes constraints on the sort
of matter couplings of $\phi$ and $\sigma$. 
From (\ref{eq:g}) follows $\Gamma_{\mathrm{RH}} \geq H_{\mathrm{reh}}$,
which combined with~(\ref{eq:j}) yields
\begin{equation}
 \frac{\Gamma_{\mathrm{DM}}}{\Gamma_{\mathrm{RH}}}
< \frac{H_0 }{H_{\mathrm{reh}}}.
\label{eq:gj}
\end{equation}
Here, if the ratio of the decay rates is determined by the mass ratio as 
\begin{equation}
  \frac{\Gamma_{\mathrm{DM}}}{\Gamma_{\mathrm{RH}}} \sim
\left( \frac{m_\sigma }{m_\phi}  \right)^n
\label{eq:G-m-n}
\end{equation}
with a positive power $n > 0$, then (\ref{eq:gj}) translates into
\begin{equation}
 \frac{m_\sigma }{m_\phi}  \lesssim
\left( \frac{H_0}{H_{\mathrm{reh}}}  \right)^{\frac{1}{n}}.
\end{equation}
This combined with $\alpha^2 < 1$ (cf. (\ref{eq:a-range})) bounds the
abundance~(\ref{eq:A2}) from above as 
\begin{equation}
 \Omega_{\mathrm{DM}} \lesssim
10^{-2} 
\left( \frac{H_{\mathrm{reh}}}{H_0} \right)^{\frac{n-4}{2n}}.
\end{equation}
Since $H_{\mathrm{reh}} / H_0 \gg 1$, this bound shows that the power
must satisfy $ n > 4$ in order to reproduce the observed dark matter
abundance. 
One particularly sees that with the axion-photon coupling $\sigma
F\tilde{F}$ alone, which gives $n = 3$, the inflaxion cannot provide all
of the dark matter.\footnote{Instead of
using the simplifying assumption~(\ref{eq:G-m-n}) where the 
dependencies on other parameters are dropped,
one can also use the full expression~(\ref{eq:sgg}) and find that, unless
$\abs{\alpha}$ is extremely close to unity, 
the coupling~$\sigma F\tilde{F}$ alone does not allow inflaxion dark
matter. However we should also note that if (\ref{eq:A}) is not
satisfied, then the abundance is no longer given by (\ref{eq:A2}) and
then it might be possible to have dark matter with axionic
couplings only.} 
A similar story holds for a derivative coupling of the axion to fermions 
of $\bar{\psi} \gamma^\mu \gamma^5 \psi \, \partial_\mu \sigma $
which gives $ n = 1$.
On the other hand, the inflaton's couplings such as
$\phi F \tilde{F}$ ($n=7$)
and $\phi \bar{\psi} i \gamma^5 \psi$ ($n=5$)
fulfill the requirement.

\begin{figure}[t]
\centering
\subfigure[$\sigma F\tilde{F} + \phi F \tilde{F}$]{%
  \includegraphics[width=.46\linewidth]{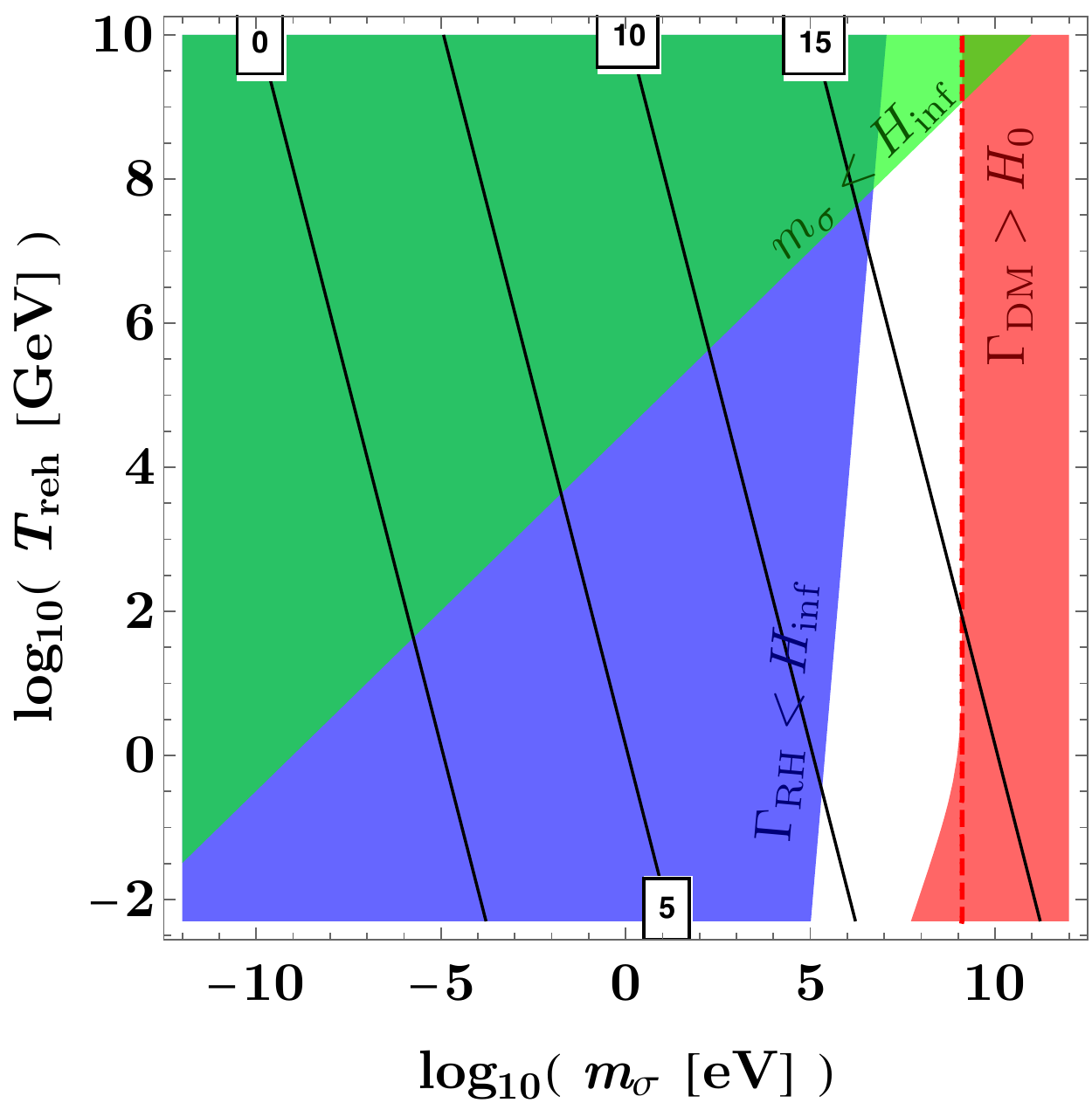}
  \label{fig:axionlike-phi2gamma}}
\, 
\subfigure[$\sigma F\tilde{F} + \phi \bar{\psi} i \gamma^5 \psi$]{%
  \includegraphics[width=.46\linewidth]{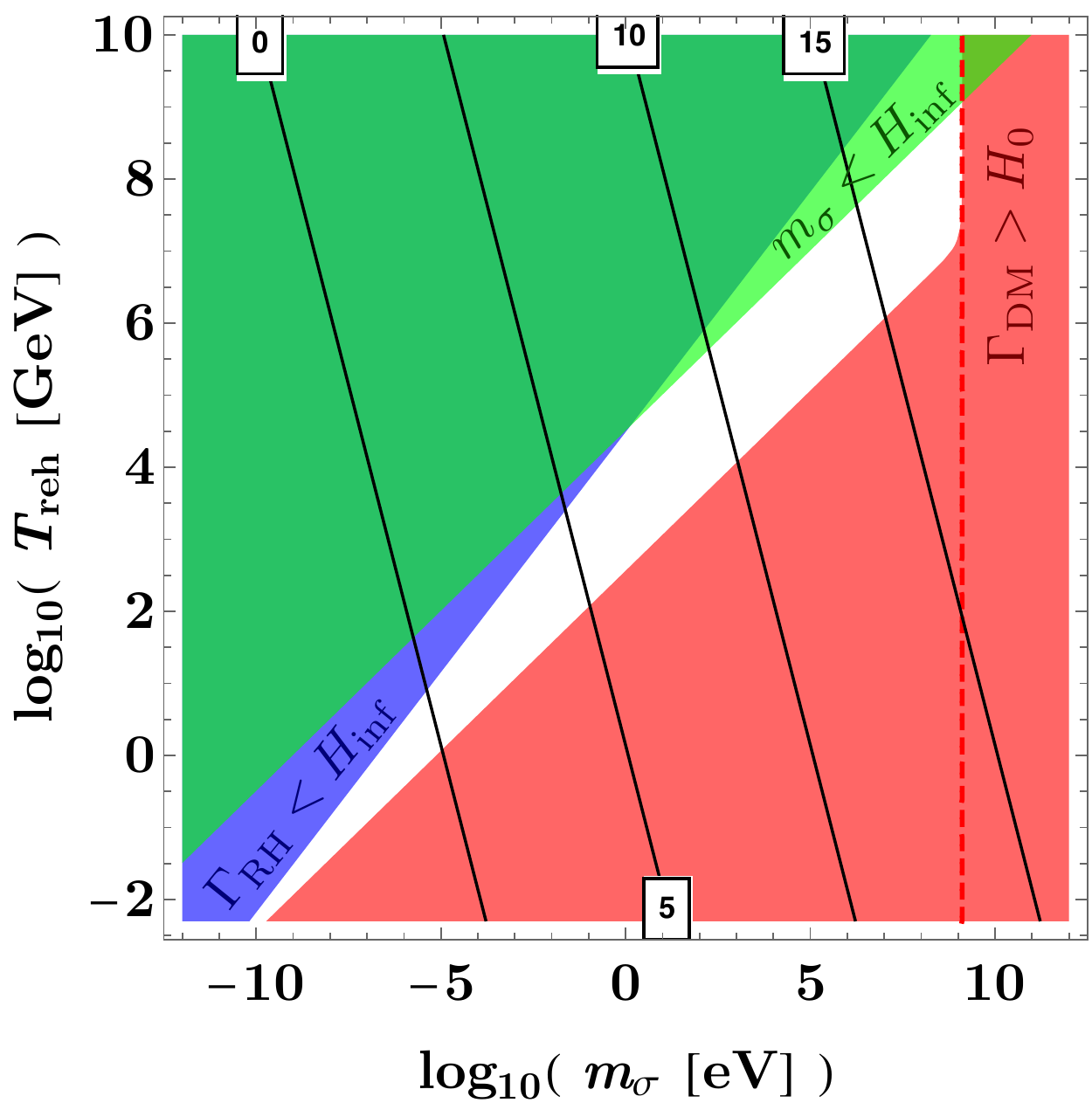}
  \label{fig:axionlike-phi2f}}
 \caption{Parameter space of axion-like fields in terms of its
 mass~$m_\sigma$ and reheating temperature~$T_{\mathrm{reh}}$. 
 Regions where the inflaxion makes up the dark matter is shown in white.
 The colored regions are excluded from the requirements of axion
 stabilization during inflation (green), dark matter stability (red),
 and instantaneous reheating (blue). The lower end of the displayed
 $T_{\mathrm{reh}}$ region is set by BBN. The axion decay constant is
 fixed to $f=10^{17}\, \mathrm{GeV}$, and the inflaton-axion mixing
 constant to $\alpha = 1/2$. 
 In the left plot the inflaton is coupled to photons with $G_{\phi
 \gamma \gamma } = 10^{-10}\, \mathrm{GeV}^{-1}$, and in the right plot
 to fermions with $g_{\phi ff} = 10^{-3}$. Black solid contours show the
 inflaton mass in terms of $\log_{10} (m_\phi / \mathrm{eV})$.
 Red dashed lines show the edge of the dark matter stability exclusion
 region if the dark matter inflaxion were allowed to decay only through
 the $\sigma F \tilde{F}$ channel.}
\label{fig:axionlike}
\end{figure}

In Figure~\ref{fig:axionlike} we show the parameter space for 
an axion-like~$\sigma$ in the $m_\sigma$ -- $T_{\mathrm{reh}}$ plane. 
Here we assumed the axion mass to be
constant throughout the cosmic evolution and thus ignored
the conditions that involve $\Lambda$.
The matter couplings are chosen in Figure~\ref{fig:axionlike-phi2gamma} as
\begin{equation}
 L_c = 
\frac{\alpha_{\gamma}}{8 \pi f}
\sigma F_{\mu \nu} \tilde{F}^{\mu \nu} 
 +  \frac{G_{\phi \gamma \gamma }}{4}
\phi F_{\mu \nu} \tilde{F}^{\mu \nu},
\label{eq:coupling-a}
\end{equation}
and in Figure~\ref{fig:axionlike-phi2f} as
\begin{equation}
 L_c = 
\frac{\alpha_{\gamma}}{8 \pi f}
\sigma F_{\mu \nu} \tilde{F}^{\mu \nu} 
+ g_{\phi ff} \, \phi \bar{\psi} i \gamma^5 \psi,
\label{eq:coupling-b}
\end{equation}
where the axion's coupling to the SM/hidden photons is taken 
to be inversely proportional to the decay constant~$f$,
with $\alpha_{\gamma}$ set equal to the electromagnetic fine
structure constant~$ 1/137$.
To simplify the analysis, we have further imposed
\begin{equation}
 \Gamma_{\mathrm{RH}} > H_{\mathrm{end}},
\label{eq:s1}
\end{equation}
so that $\varphi_{\mathrm{RH}}$ instantaneously decays at the end of
inflation and reheats the universe, rendering $H_{\mathrm{reh}} =
H_{\mathrm{end}}$
(see also discussions in Footnote~\ref{fn:1}). 
The other parameters are fixed to
$f=10^{17}\, \mathrm{GeV}$, $\alpha = 1/2$, $C = 10$,
and the coupling constant in Figure~\ref{fig:axionlike-phi2gamma}
as $G_{\phi \gamma \gamma } = 10^{-10}\, \mathrm{GeV}^{-1}$,
and in Figure~\ref{fig:axionlike-phi2f} as $g_{\phi ff} = 10^{-3}$.
The inflaton mass~$m_\phi$ is fixed by requiring $\varphi_{\mathrm{DM}}$
to make up the entire dark matter abundance, 
i.e. $\Omega_{\mathrm{DM}} \approx 0.3$, 
and its value is shown by the black solid contours labeled with
$\log_{10} (m_\phi / \mathrm{eV})$.
For the chosen set of parameters, 
the conditions that most strongly restrict the parameter space
are axion stabilization during inflation~(\ref{eq:c}) which excludes the
green regions in the plots,
dark matter stability~(\ref{eq:j}) excluding the red regions,
instantaneous reheating~(\ref{eq:s1}) excluding the blue regions,
as well as the BBN constraint~(\ref{eq:h}) setting the lower bound of the 
displayed $T_{\mathrm{reh}}$-range.\footnote{In the plots we have
ignored the time variation of the Hubble rate during inflation, thus
$H_{\mathrm{end}}$ is written as~$H_{\mathrm{inf}}$.}
The allowed windows are shown in white. 
We note that $m_\sigma$ is approximately the same as the
mass~$m_{\mathrm{DM}}$ of the inflaxion dark matter field,
and since we are supposing instantaneous reheating,
$T_{\mathrm{reh}}$ is the same as the inflationary energy scale
up to a factor that depends on~$g_*$.

The conditions on the decay rates obviously depend on the details of the
matter couplings.  
As can be seen from the blue regions excluded by
(\ref{eq:s1}) taking different shapes in the two plots,
the reheaton~$\varphi_{\mathrm{RH}}$ decays predominantly through the
inflaton's couplings in the allowed windows.
For the dark matter~$\varphi_{\mathrm{DM}}$, both of the inflaton and
axion couplings can provide the main decay channel. 
The red dashed lines in the plots 
show where the condition~(\ref{eq:j}) would be saturated if
$\varphi_{\mathrm{DM}}$ were allowed to decay only through the axion's
$\sigma F \tilde{F}$ coupling; one sees that at
higher~$T_{\mathrm{reh}}$ the decay of~$\varphi_{\mathrm{DM}}$ is
governed by this axion coupling, while at lower~$T_{\mathrm{reh}}$ by
the inflaton couplings. 
Different inflaton couplings open different windows inside the
triangular area between the green region and the red dashed line. 
In particular in Figure~\ref{fig:axionlike-phi2f}, if the fermion mass
lies between those of the dark matter and reheaton as 
$m_{\mathrm{DM}} < 2 m_{\psi} < m_{\mathrm{RH}}$, 
then the decay of the dark matter into fermions would
be kinetically forbidden and hence the right edge of the allowed window
would be extended to the red dashed line. 
In both plots, as the coupling constants $G_{\phi \gamma \gamma}$ and
$g_{\phi ff}$ are increased, the regions sandwiched between the blue and
red shift or expand to smaller $m_\sigma$ values. 
We also note that the allowed windows become larger as $f$ is increased,
and as $\abs{\alpha}$ is taken closer to unity.

With the conventional vacuum misalignment scenario, axions with masses
and decay constants as shown here would overdominate the
universe, unless the initial angle takes extremely tiny values. 
This can be readily checked from~(\ref{eq:axion-like-ab});
given a vacuum misalignment angle of $\abs{\theta_\star} \sim 1$, 
to make up the dark matter abundance with $f = 10^{17}\, \mathrm{GeV}$
would require an ultralight mass $m_\sigma \sim 10^{-22}\, \mathrm{eV}$,
which on the other hand is in tension with the small-scale observations
of the universe. 
However with the inflaxion mechanism, the observed dark
matter abundance can be achieved with much larger axion masses,
even under a number of conditions we have imposed only for the 
purpose of simplifying the analysis.

If it is the SM photon that the axion and inflaton are coupled to, 
it is natural to ask
how the existing~\cite{Tanabashi:2018oca} and forecasted~\cite{Agrawal:2017cmd} 
experimental constraints on such couplings 
compare to those in our Figure~\ref{fig:axionlike}. 
With $f = 10^{17}$ GeV, the axion coupling is very small and hard to test, but in the mass
region $m_\sigma  \lesssim 10^{-7}\, {\rm eV}$ it could almost be within
reach of the upcoming 
ABRACADABRA experiment~\cite{Kahn:2016aff,Ouellet:2019tlz}. 
The inflaton coupling,
$G_{\phi \gamma \gamma } = 10^{-10}\, \mathrm{GeV}^{-1}$,
could saturate the upper
bounds from globular cluster stars~\cite{Ayala:2014pea} and the 
CAST helioscope~\cite{Anastassopoulos:2017ftl},
depending on the mass of~$\varphi_{\mathrm{RH}}$.
However within the allowed window shown in
Figure~\ref{fig:axionlike-phi2gamma},
it is out of reach of current experiments~\cite{Tanabashi:2018oca}.
The effective coupling for $\varphi_{\mathrm{DM}}$ 
obtained through the inflaton is
much smaller than~$G_{\phi \gamma \gamma}$ (cf.~(\ref{eq:318})).

In the plots we have considered an axion-like field whose mass stays
constant during the cosmic evolution.
When $\sigma$ is the QCD axion, the condition~(\ref{eq:A})
requires the reheating temperature to be below the QCD scale, which
might pose a challenge for realizing an observationally viable
cosmological scenario.
(The same is true for axion-like fields
lying in regions in the plots close to the lower edges.)
Under this constraint, to open a QCD inflaxion window at $f \gtrsim
10^{9}\, \mathrm{GeV}$ with the reheaton decaying into the SM photons
as in~(\ref{eq:coupling-a}) requires a very large
coupling~$G_{\phi \gamma \gamma}$, which would be experimentally excluded.
On the other hand, a reheaton decaying into non-SM particles at such
low reheating temperatures could also cause tension with observations, 
including the BBN and CMB constraints on extra radiation.
As a proof of concept, 
we present the parameter space for QCD inflaxions 
with low-scale reheating in Appendix~\ref{app:QCDinflaxion}. 
However, since (\ref{eq:A}) was adopted only to simplify the analysis,
it would be important to explore the possibilities beyond this
restriction, which we leave for future work.

\section{Conclusions}
\label{sec:conc}

We have proposed a dynamical production mechanism for axion dark matter. 
The three key ingredients of the construction are low-scale inflation, 
a kinetic mixing between the axion and inflaton, and an inflaton mass
after inflation that is larger than the axion mass. 
The axion during inflation is driven to a field
point slightly displaced from the potential minimum, which enables 
the axion combined with the inflaton 
to make up the observed dark matter abundance.
This opens up new parameter windows for the axion,
enabling it to serve as dark matter with masses and decay constants
that would lead to overproduction
in the conventional vacuum misalignment scenario.
We showed in particular that axion-like particles
with decay constants of $f \sim 10^{17}\, \mathrm{GeV}$
can make up dark matter without having extremely tiny initial
misalignment angles, or ultralight masses that would
spoil the small-scale structure of the universe.
Although in this paper we did not investigate the full range of
possibilities for the QCD axion, our mechanism can in principle also
work for QCD axions with similarly large decay constants.
The dark matter abundance produced by the inflaxion mechanism 
is uniquely determined by the Lagrangian parameters, unlike the case for
the vacuum misalignment.
Also due to the fact that inflaxion dark matter is free of isocurvature
perturbations, it escapes from conventional cosmological constraints on
axions.

An important feature of the inflaxion mechanism is that 
the kinetic mixing, together with the rolling of the inflaton,
rotates the diagonalized basis in the field space.
This leads to different combinations of the inflaton
and the axion fields, i.e. the inflaxions,
to each generate the cosmological perturbations, reheat the universe,
and serve as dark matter.  
The kinetic mixing also induces inflaxion dark matter to 
interact with other particles through both the axion couplings and
the inflaton couplings. 
This in particular relates the reheating temperature of the
universe with the lifetime of inflaxion dark matter.
Moreover, the dark matter lifetime can be rather short---not much longer
than the age of the universe---even if the axion decay constant is large. 
These facts about the inflaxion couplings offer the possibility of
producing smoking-gun signatures for distinguishing the inflaxion
mechanism from other axion production scenarios such as the vacuum
misalignment or emission from cosmic strings.  
For instance, the case where inflaxion dark matter decays with a
lifetime close to the age of the universe can give rise to a number of
observable signals, and may even relax the 
recent tensions in the measurements of the matter fluctuation
amplitude of the universe~\cite{Enqvist:2015ara}. 
Furthermore, parts of the axion window that our mechanism opens up can be
probed by upcoming experiments such as ABRACADABRA. 
We also remark that the matter couplings the diagonal fields pick up
through the axion could give rise to parity-violating signatures in the sky~\cite{Lue:1998mq}.
Moreover, if the duration of inflation is minimal and thus
at the time when the CMB scales exit the horizon
the axion is still oscillating with a non-negligible amplitude,
then its mixing with the inflaton could give rise to oscillatory features
in the curvature perturbation spectrum on observable scales.

The basic idea of the mechanism is to dilute away the axion density
with low-scale inflation, while slightly exciting the axion from
its post-inflationary vacuum state. 
The inflaxion achieves this through a derivative coupling with
the slow-rolling inflaton. One may also construct variants with, 
e.g., a shift-symmetric Gauss-Bonnet coupling~$\sigma  
\mathcal{G}_{\mathrm{GB}}$ (which can also allow axions to induce
baryogenesis~\cite{DeSimone:2016bok}), 
or $\sigma F\tilde{F}$, given that there is a background of helical
electromagnetic fields which can in principle be produced as
in~\cite{Anber:2006xt,Durrer:2010mq,Bastero-Gil:2018uel}. 
An alternative possibility would be to generate an additional small
potential, possibly from some hidden gauge interactions that become
strong only during inflation 
(see e.g.~\cite{Dvali:1995ce,Dine:2004cq,Takahashi:2015waa}).

We stress that some of the requirements for the inflaxions
discussed in this paper, such as the axion potential to stay constant
during reheating, were introduced merely for simplifying the analyses. 
In particular for the QCD axion, this implied a very low reheating
temperature. 
However by lifting such simplifying conditions, other scenarios can also
arise within the framework, such as a temporal vanishing of the axion
potential upon reheating. Another possibility we did not study is a
non-perturbative decay of the reheaton. It will be interesting to
systematically study the range of possibilities to fully explore the
axion parameter space capable of explaining the dark matter of our
universe. 
Another important direction for further study is to realize inflaxions
within an ultraviolet-complete framework. 
The inflaton-axion kinetic mixing can arise, for instance, 
in the low-energy limit of string compactifications via
non-perturbative corrections to the K\"{a}hler
potential
in a similar fashion to~\cite{Cicoli:2017zbx}.
The inflaxion mechanism could also provide a new cosmological picture of
the axiverse, in which most axions settle down to the vacuum
during low-scale inflation, except for those coupled to the
inflaton and thence get excited to make up the dark matter.
This picture may offer a way to avoid the cosmological moduli problem
in the axiverse. 
We leave these topics for future investigation.

\section*{Acknowledgments}

We thank Toyokazu Sekiguchi for useful comments on a draft, and
Enrico Barausse, Sebastian Cespedes, Michele Cicoli, Djuna Croon, Gilly
Elor, Diptimoy Ghosh, Marco Gorghetto, Veronica Guidetti, Ann Nelson, and Giovanni Villadoro for helpful discussions.

%\newpage

\appendix

\section{Diagonal Basis}
\label{sec:diagonal}

In this appendix we diagonalize the kinetic and mass terms in the
Lagrangian~(\ref{eq:Lagrangian}). 
Here we expand the effective potential~$V(\phi)$ around some field
value~$\phi_*$ up to quadratic order, 
\begin{equation}
 V(\phi) = V_* + V_*' (\phi - \phi_*) + \frac{1}{2} V_*''
  (\phi - \phi_*)^2 + \mathcal{O}(\phi - \phi_*)^3,
\label{eq:quad}
\end{equation}
where $V_* = V(\phi_*)$, $V_*' = dV(\phi)/d\phi |_{\phi = \phi_*}$, etc.,
and we note in particular that the effective mass squared~$V_*''$ can be
either positive or negative. 
(In the main text the axion mass squared is considered to be $m_\sigma^2
> 0$, however the discussion in this appendix also applies to tachyonic
masses, i.e. $m_\sigma^2 < 0$. 
Below we ignore any temperature dependence of the axion mass.)
We can complete the square by the shift 
\begin{equation}
 \tilde{\phi} = \phi - \phi_* + \frac{V_*'}{V_*''}.
\end{equation}
The kinetic and potential terms up to quadratic order 
can then be diagonalized by the following field redefinitions, 
\begin{equation}
 \varphi_{\pm} = 
\left(\frac{X_{\pm} +  \beta^2}{ X_{\pm}^2 + \alpha^2 \beta^2 }\right)^{1/2}
\left(  X_{\pm} \, \sigma + \alpha \, \tilde{\phi} \right),
\label{varphi_pm}
\end{equation}
where
\begin{equation}
 \beta^2 = \frac{m_\sigma^2}{V_*''},
\quad
 X_{\pm} =  \frac{ 1 -\beta^2 \pm \sqrt{ (1 - \beta^2)^2 + 4 \alpha^2
 \beta^2 }}{2}.
\end{equation}
One can check that the coefficients of $\sigma$ and $\tilde{\phi}$
in (\ref{varphi_pm}) are real, hence $\varphi_\pm$ are guaranteed to be
real fields.
The Lagrangian~(\ref{eq:Lagrangian}) is rewritten in terms of these
fields as
\begin{equation}
 \frac{\mathcal{L}}{\sqrt{-g} } = 
\sum_{i=\pm} \left(
-\frac{1}{2} g^{\mu \nu} \partial_\mu \varphi_i \, \partial_\nu \varphi_i
- \frac{1}{2} m_i^2 \varphi_i^2 
\right)
- V_* + \frac{(V_*')^2}{2 V_*''}
- \mathcal{O}(\phi - \phi_*)^3 + L_c[\sigma, \phi, \Psi],
\label{S-diag}
\end{equation}
with the mass squareds given by
\begin{equation}
 m_{\pm}^2 = \frac{1 -  X_{\pm} }{1 - \alpha^2} \, V_*''.
\end{equation}
Here, note that $m_{\pm}^2$ can be negative. 

In the presence of a hierarchy between the masses such that
\begin{equation}
 \abs{m_{\sigma}^2} \gg \abs{V_*''},
\label{eq:sggp}
\end{equation}
the expression~(\ref{varphi_pm}) can be simplified.
By expanding each of the coefficients of $\sigma$ and $\tilde{\phi}$ 
to leading order in $V_*'' / m_\sigma^2$, one obtains\footnote{We compute the leading order
expressions for the coefficients, since $\sigma$ and $\tilde{\phi}$
themselves can also take hierarchical field values that depend on their
masses.}
\begin{equation}
\varphi_{\mathrm{L}} \simeq
\alpha  \sigma + \tilde{\phi},
\quad
\varphi_{\mathrm{H}} \simeq
\sqrt{1 - \alpha^2}
\left(
 \sigma  -\frac{\alpha V_*''}{m_{\sigma}^2} 
 \tilde{\phi} 
\right),
\label{varphi_HL}
\end{equation}
up to overall signs which do not affect the physics. 
Here $\varphi_{\mathrm{L}}$ and $\varphi_{\mathrm{H}}$ denote respectively 
the lighter and heavier of the two fields~$\varphi_{\pm}$, 
whose masses at the leading order are
\begin{equation}
 m_{\mathrm{L}}^2 \simeq  V_*'',
\quad
 m_{\mathrm{H}}^2 \simeq \frac{ m_{\sigma}^2}{1-\alpha^2}.
\label{eq:m_HL}
\end{equation}

The expressions in the opposite hierarchical regime $\abs{m_{\sigma}^2} \ll \abs{V_*''}$ can similarly be obtained by exchanging
$(\sigma, m_{\sigma}^2) \leftrightarrow (\tilde{\phi}, V_*'') $.

\section{Misalignment of Stabilized Axion}
\label{sec:displacement}

\subsection{Generic Cosmological Background}

Let us analyze the field displacement of a stabilized axion~$\sigma$ 
using the diagonal basis introduced in Appendix~\ref{sec:diagonal}.
In the following discussions, the $\mathcal{O}(\phi - \phi_*)^3$ terms 
in the potential expansion~(\ref{eq:quad}) and the 
matter coupling~$ L_c[\sigma, \phi, \Psi]$ are neglected.

We consider a flat FRW universe~(\ref{eq:FRW}), 
and suppose the axion mass to be larger than
that of~$\phi$ as well as the Hubble rate, 
\begin{equation}
 m_\sigma^2 \gg H^2, \, \abs{V_*''}. 
\label{eq:B1}
\end{equation}
The approximate expressions for the diagonal fields for such a case
are given in~(\ref{varphi_HL}). 
From (\ref{eq:m_HL}) combined with (\ref{eq:a-range}),
one sees that the heavier field~$\varphi_{\mathrm{H}}$ also possesses a
super-Hubble mass, i.e. $m_{\mathrm{H}}^2 \gg H^2$.
Hence $\varphi_{\mathrm{H}}$ undergoes a damped oscillation about its
origin, and thus eventually gets stabilized at 
$\varphi_{\mathrm{H}} = 0$.
In the $(\sigma, \phi)$ basis, this corresponds to 
\begin{equation}
  \sigma  \simeq \frac{\alpha V_*''}{m_{\sigma}^2} \,
   \tilde{\phi}  \simeq  \frac{\alpha V'(\phi)}{m_{\sigma}^2},
\label{maru-i}
\end{equation}
which shows that the axion field~$\sigma$ is displaced from its origin.

\subsection{Slow-Roll Inflation}

We now consider the lighter diagonal field~$\varphi_{\mathrm{L}}$ to
be a slow-rolling inflaton. 
Let us combine its mass term and the constant offset in~(\ref{S-diag}) into
\begin{equation}
 U(\varphi_{\mathrm{L}}) = V_* -
\frac{(V_*')^2}{2 V_*''}
+  \frac{1}{2} m_{\mathrm{L}}^2 \varphi_{\mathrm{L}}^2.
\end{equation}
Then one can check that once the axion
is stabilized as~(\ref{maru-i}),
the effective potentials in the two bases
$U(\varphi_{\mathrm{L}})$ and $V(\phi)$ agree as
\begin{equation}
U(\varphi_{\mathrm{L}}) \simeq V(\phi), \quad
U'(\varphi_{\mathrm{L}}) \simeq V'(\phi), \quad
U''(\varphi_{\mathrm{L}}) \simeq V''(\phi),
\label{eq:UV}
\end{equation}
where the derivatives of $U$ and $V$ are understood to be in terms
of~$\varphi_{\mathrm{L}}$ and $\phi$, respectively.
Hence the values of the slow-roll parameters defined in the
diagonal basis
\begin{equation}
 \epsilon = \frac{M_p^2}{2} \left( 
\frac{U'(\varphi_{\mathrm{L}})}{U(\varphi_{\mathrm{L}})}
 \right)^2,
\quad
\eta =  M_p^2 \frac{U''(\varphi_{\mathrm{L}})}{U(\varphi_{\mathrm{L}})},
\end{equation}
agree with those defined similarly in terms of $V(\phi)$. 
When these parameters are smaller than unity,\footnote{With
$\abs{\eta} \ll 1$, the mass hierarchy~(\ref{eq:B1}) becomes
$m_\sigma^2 \gg H^2 \gg \abs{V_*''}$.}
$\varphi_{\mathrm{L}}$ follows the slow-roll attractor,
\begin{equation}
 3 M_p^2 H^2 \simeq U(\varphi_{\mathrm{L}}), \quad
 3 H \dot{\varphi_{\mathrm{L}}} \simeq -U'(\varphi_{\mathrm{L}}).
\end{equation}
One can further check that from these approximations follow
\begin{equation}
 3 M_p^2 H^2 \simeq V(\phi ), \quad
 3 H \dot{\phi} \simeq -V' (\phi).
\end{equation}
Hence we see that with the axion stabilized, 
the slow-roll conditions and approximations for $\phi$ take 
the same forms as those for a canonical single-field inflaton.
Moreover, during slow-roll inflation, the axion
displacement~(\ref{maru-i}) can also be written as
\begin{equation}
  \sigma \simeq - \frac{3 \alpha H\dot{\phi}}{m_{\sigma}^2}.
\end{equation}

The end of inflation, i.e. when the sign of $\ddot{a}$ changes from
positive to negative, corresponds to the time when $-\dot{H}/H^2=1$.
Given that $\varphi_{\mathrm{H}}$ continues to be stabilized at~$0$, 
this happens when the inflaton potential and kinetic energies are
related by 
\begin{equation}
 U(\varphi_{\mathrm{L}}) = \dot{\varphi_{\mathrm{L}}}^2 = 2 M_p^2 H^2.
\end{equation}
Using~(\ref{eq:UV}), the end of inflation can also be described 
in terms of~$\phi$ as the time of
\begin{equation}
 V(\phi) \simeq 2 M_p^2 H^2.
\label{eq:end-V}
\end{equation}
This relation, however, can receive corrections if for instance the 
higher order terms in the potential expansion~(\ref{eq:quad}) becomes
important or the inflaton mass~$\abs{V''}$ becomes
larger than~$m_\sigma^2$ towards the end of inflation.

\section{QCD Inflaxion}
\label{app:QCDinflaxion}

Before turning to the QCD inflaxion, we briefly review the conventional
story of QCD axion dark matter. 
If the Peccei-Quinn~\cite{Peccei:1977hh} symmetry breaking happens before or during the inflationary
epoch and is not restored afterwards, then 
the present-day abundance of the QCD axion arising from
a vacuum misalignment is~\cite{Turner:1985si}
\begin{equation}
 \Omega_\sigma  \sim 0.1 \times \theta_\star^2 
\left( \frac{f}{10^{12} \, \mathrm{GeV}} \right)^{7/6}.
\label{eq:QCDaxion-ab}
\end{equation}
An axion with an initial misalignment angle $\abs{\theta_\star} \sim 1$
makes up the observed dark matter abundance if the decay
constant is $f \sim 10^{12}\, \mathrm{GeV}$.
However for $f \gg 10^{12}\, \mathrm{GeV}$, 
the axion would overdominate the universe 
unless the misalignment angle is tuned to lie near zero,
i.e. $\theta_\star^2 \ll 1$ 
(``anthropic axion window''~\cite{Tegmark:2005dy}).
On the other hand for $f \ll 10^{12}\, \mathrm{GeV}$, the axion could
make up the entire dark matter only if 
$\theta_\star \simeq \pm \pi$ which renders the axion abundance 
larger than the expression~(\ref{eq:QCDaxion-ab}) due
to anharmonic effects~\cite{Turner:1985si,Bae:2008ue}. 
However in this case the axion isocurvature perturbations would also be
enhanced and thus such anharmonic axion dark matter is severely
constrained by CMB 
measurements~\cite{Lyth:1991ub,Strobl:1994wk,Kobayashi:2013nva}. 

If, on the other hand, the symmetry breaks after inflation
or is restored upon reheating, then axionic domain walls would form and
overdominate the universe. 
The walls however can be made unstable by
setting the number of degenerate vacua of the axion potential (the
so-called domain wall number) to $N_{\mathrm{W}} =
1$~\cite{Vilenkin:1982ks,Sikivie:1982qv,Linde:1990yj,Lyth:1992tx};
alternatively, one can have $N_{\mathrm{W}} > 1$ and gauge the
$Z_{N_{\mathrm{W}}}$~symmetry so that all vacua
are gauge equivalent~\cite{DiLuzio:2017tjx}, 
or introduce further breaking of the axion's shift symmetry to generate
an energy bias between the vacua~\cite{Zeldovich:1974uw,Gelmini:1988sf,Coulson:1995nv,Larsson:1996sp}. 
In these cases, axions are radiated from the axionic domain
walls and cosmic strings and thus can in principle make up the dark
matter.
However the amount of axion dark matter produced is still uncertain;
see~\cite{Gorghetto:2018myk,Kawasaki:2018bzv}
for recent works on axion strings.

\begin{figure}[t]
  \begin{center}
  \begin{center}
  \includegraphics[width=.48\linewidth]{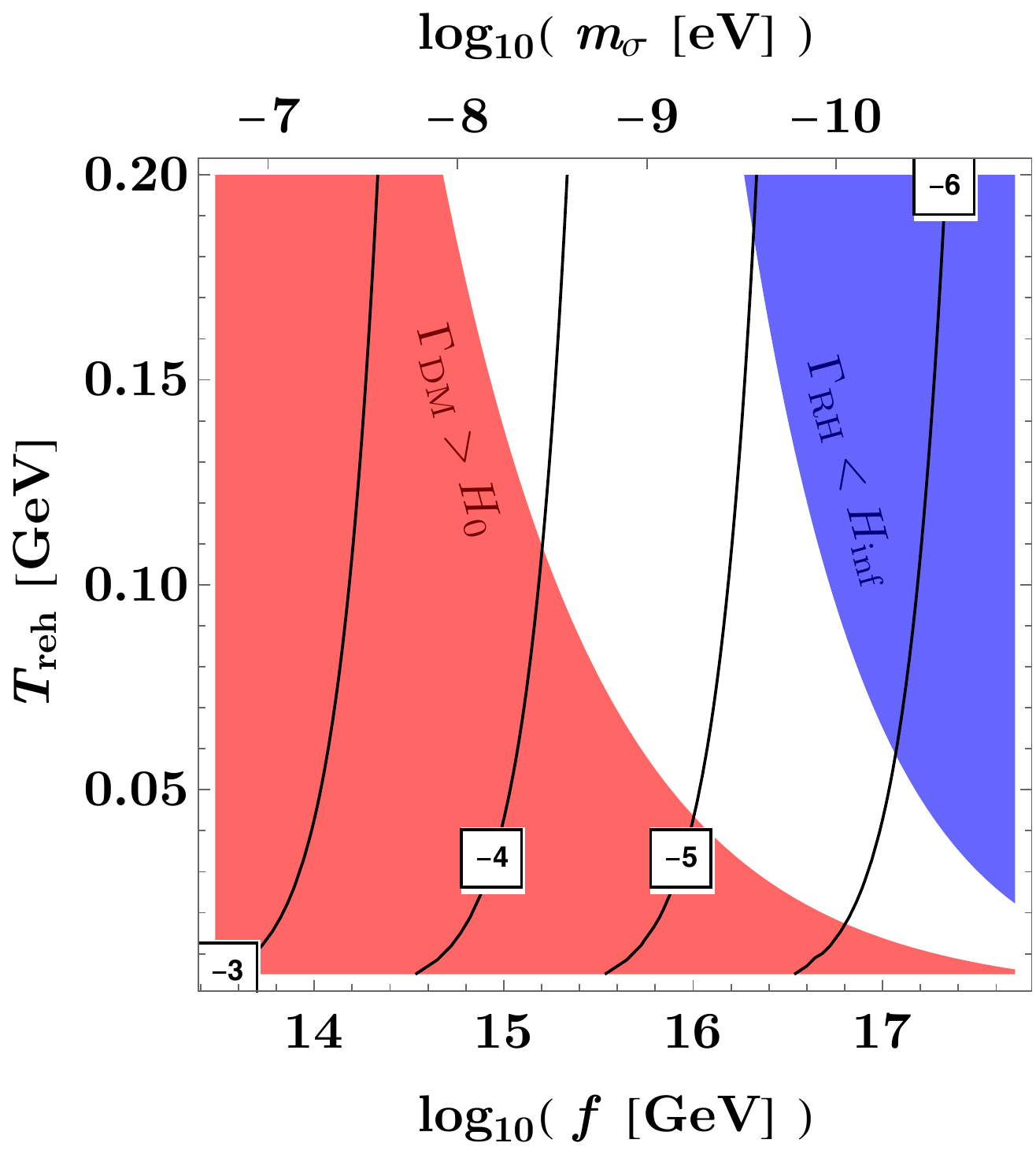}
  \end{center}
  \caption{Parameter space of the QCD axion in terms of its decay
   constant~$f$, mass~$m_\sigma$, and reheating
   temperature~$T_{\mathrm{reh}}$. 
   To simplify the analyses, $T_{\mathrm{reh}}$ is taken to be below the
   QCD scale; with such extremely low reheating
   temperatures, the axion window shown here should be considered
   as a proof of concept (see the main text for discussions).
   Regions where the QCD inflaxion can make
   up the dark matter is shown in white. The colored regions are
   excluded from the requirements of dark matter stability (red) and
   instantaneous reheating (blue). The lower end of the displayed
   $T_{\mathrm{reh}}$ region is set by BBN.
   The inflaton-axion mixing constant is fixed to $\alpha =
   1/2$, and the inflaton is coupled to fermions with $g_{\phi ff}  =
   5 \times 10^{-3}$. Black solid contours show the inflaton mass in
   terms of $\log_{10} (m_\phi / \mathrm{eV})$.}
  \label{fig:QCDaxion}
  \end{center}
\end{figure}

The inflaxion mechanism for the QCD axion is subject to conditions
involving the QCD scale.
In particular, the condition~(\ref{eq:A}) adopted in the main text
restricts the reheating temperature to be below the QCD scale, 
which poses a challenge for realizing an observationally viable
cosmology with the inflaxions.
Although (\ref{eq:A}) is merely introduced to simplify the analyses and
not a necessary condition for the inflaxion framework, 
below we demonstrate as a proof of concept that all the conditions
listed at the beginning of Section~\ref{sec:parameter} can in principle
be met for the QCD axion. 
In the parameter region we will show in Figure~\ref{fig:QCDaxion},
the reheaton mass is much lighter than eV. Under this condition, 
one example of the accompanying reheating scenario 
would be that the reheaton decays into very light non-SM 
fermions, which couple to SM particles and thus 
thermalize the visible sector above the BBN temperature. 
Here, in order to avoid the observational constraints
on extra radiation~\cite{Aghanim:2018eyx},
the non-SM fermion is assumed to have a time-varying
mass, arising for instance from a coupling to a rolling scalar field,
which quickly renders the fermion non-relativistic after reheating
so that they do not contribute to $\Delta N_{\rm eff}$.

The parameter space is shown in Figure~\ref{fig:QCDaxion} 
in terms of the axion decay constant~$f$ 
and the reheating temperature~$T_{\mathrm{reh}}$.
The axion mass~$m_\sigma$ is a function of~$f$ as shown in
(\ref{eq:QCDmass}).  
We used a linear scale for~$T_{\mathrm{reh}}$,
and the lower end of the displayed $T_{\mathrm{reh}}$-range is set by
the BBN constraint~(\ref{eq:h}),
while the upper end is set to~$\Lambda_{\mathrm{QCD}}$
from the simplifying condition~(\ref{eq:A}).
We adopted the parameter values $\alpha = 1/2$ and $C= 10$, 
required instantaneous reheating~(\ref{eq:s1}),
and fixed the inflaton mass such that $\Omega_{\mathrm{DM}} \approx 0.3$;
the black solid contours are labeled with $\log_{10} (m_\phi /
\mathrm{eV})$. 
Moreover, we have considered couplings as in (\ref{eq:coupling-b})
between the axion and photons,\footnote{For the QCD axion-photon
coupling, there is a further multiplicative factor of order unity which
is set by the electromagnetic and color
anomalies~\cite{diCortona:2015ldu}. However the detailed value of this
model-dependent factor does not affect the analysis here since the decay
channels are dominated by the inflaton's coupling.}
as well as the inflaton and fermions with 
$g_{\phi ff} = 5 \times 10^{-3}$;
the latter interaction dominantly determines the decay rates of both
$\varphi_{\mathrm{RH}}$ and $\varphi_{\mathrm{DM}}$
in the displayed parameter range.
The conditions that most strongly restrict the parameter space 
besides (\ref{eq:h}) and (\ref{eq:A}) are the
dark matter stability~(\ref{eq:j}) excluding the red region,
and instantaneous reheating~(\ref{eq:s1}) excluding the blue region.
The white region denotes the allowed window.
As the coupling~$g_{\phi ff}$ is decreased,
the position of the allowed window shifts towards smaller $f$~values,
hence larger axion and inflaton masses;
the value of~$g_{\phi ff}$ in the plot is chosen such that the allowed
window is centered at $f \sim 10^{16}\, \mathrm{GeV}$. 
We should also remark that the range of
$ 3 \times 10^{17}\, \mathrm{GeV} < f < 1 \times 10^{19}\,
\mathrm{GeV}$, 
which includes the region close to the right edge of the plot,
is disfavored by black hole superradiance
limits~\cite{Arvanitaki:2014wva,Brito:2017wnc}.

If the inflaton is coupled instead to photons as
in~(\ref{eq:coupling-a}), 
then a very large coupling~$G_{\phi ff}$
is required in order to open a QCD axion window at 
$f \gtrsim 10^{9}\, \mathrm{GeV}$,
which would be experimentally excluded.
However this statement assumes a perturbative decay of the reheaton, 
and the picture may differ when one takes into account the
possibility of non-perturbative decay (see e.g.~\cite{Adshead:2015pva, Cuissa:2018oiw}). 
We leave a detailed study of the QCD inflaxion cosmology to future work.

%\clearpage

\bibliographystyle{JHEP}
\bibliography{axion}

\end{document}